\documentclass{Amarante_aastex}          
\usepackage{Amarante_spr-astr-addons}    
\usepackage{url}

\RequirePackage{color}
\usepackage{graphicx}	
\usepackage{amsmath,amsfonts,amssymb,amsthm, bm}

\usepackage{booktabs}
\usepackage{Amarante_physics}
\usepackage[flushleft]{threeparttable}

\def\orcidname{ORCID IDS}
\newenvironment{orcid}[1][\orcidname]{\begin{@conflict}[#1]}{\end{@conflict}}

\usepackage{hyperref}
\hypersetup{colorlinks=true, linkcolor=blue, citecolor=blue, filecolor=blue, pagecolor=blue, urlcolor=blue}

\newcommand{\specialcell}[2][c]{%
  \begin{tabular}[#1]{@{}c@{}}#2\end{tabular}}

\makeatletter
\def\@seccntformat#1{\@ifundefined{#1@cntformat}%
   {\csname the#1\endcsname\quad}  
   {\csname #1@cntformat\endcsname}
}
\let\oldappendix\appendix 
\renewcommand\appendix{%
    \oldappendix
    \newcommand{\section@cntformat}{\appendixname~\thesection:\quad}
}
\makeatother

\begin{document}
%
\title{The fate of particles in the dynamical environment around Kuiper Belt object (486958) Arrokoth}

\shorttitle{The fate of particles around KBO (486958) Arrokoth}
\shortauthors{Amarante et al.}

\author{A. Amarante$^1$
$\bullet$ 
O. C. Winter$^2$}
\affil{$^1$\href{https://orcid.org/0000-0002-9448-141X}{\includegraphics[scale=0.5]{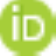}} Federal Institute of Education, Science and Technology of S\~ao Paulo - IFSP, S\~ao Jos\'e dos Campos, S\~ao Paulo, Brazil}
\affil{$^2$\href{https://orcid.org/0000-0002-4901-3289}{\includegraphics[scale=0.5]{orcid_16x16.pdf}} Grupo de Din\^amica Orbital e Planetologia, S\~ao Paulo State University - UNESP, Guaratinguet\'a, S\~ao Paulo, Brazil}


\altaffiltext{1}{andre.amarante@unesp.br (AA)}
\altaffiltext{2}{othon.winter@unesp.br (OCW)}

\begin{abstract}
The contact binary Kuiper Belt object (486958) Arrokoth, targeted by \textit{New Horizons} mission, has a unique slope pattern, which is a result of its irregular bilobate surface shape and high spin period.
Thus, some peculiar topographic regions on its surface are predisposed to lose or accumulate material, as a long circular depression feature, an impact crater called Maryland, on its small lobe.
The equilibrium points of Arrokoth are also directly related to the structure of the environment near these surface features.
In this work, we performed numerical simulations around Arrokoth to explore the fate of particles close to equilibrium points and their dynamical connection with its surface features.
Our results suggest that most of these particles in a ring inside the Arrokoth's rotational Roche lobe fall near the equatorial region of the Maryland impact crater or close to the Bright spots area on the large lobe. 
Also, particles in a spherical cloud orbiting Arrokoth accumulate preferentially near low-mid-latitudes regions close to the longitudes of Maryland crater and Bright spots area. In contrast, a few particles will fall in regions diametrically opposite to them, as in the LL\_Term boundary on the large lobe. High-latitudes are those more empty of impacts, as in polar sites.
In addition, particles larger than a couple of microns are not significantly perturbed by solar radiation pressure in the environment around Arrokoth.
\end{abstract}


\section{Introduction}
\label{sec:intro}
An overview of the initial results from the \textit{New Horizons} space probe's close-approach reconnaissance, in early 2019, revealed that the cold classical Kuiper Belt object (KBO) Arrokoth has a peculiar surface shape. It consists of a contact binary structure merged in a thin `neck' by two lobes (large and small) containing smoothed surfaces, hexagonal patches, and some regions that do have multiple morphologic indicators suggesting they are impact craters \citep{Stern2019}. Most of these potential impact craters on the surface of Arrokoth are likely to have a diameter of $<\sim$1\,km, except for a peculiar feature. This peculiar surface characteristic, unofficially named `Maryland' (MD) by the \textit{New Horizons}' mission team, is a circular depression enclosed by high regions. It is the most extensive observed feature of the Arrokoth located on the surface of its small lobe. This feature is likely to be a crater with a diameter of $\sim$7\,km \citep{Stern2019}. These topographic characteristics show that the Arrokoth's surface is only modestly cratered with a relatively benign dynamic collisional environment \citep{Singer2019}. Besides, there are no apparent craters between 1 and 7 km in diameter on its surface \citep{Spencer2020,Grundy2020}. Therefore, Arrokoth provides a record of the process of forming planetesimals. The first generation of gravitationally bound bodies that have been minimally altered by subsequent heating and impactor bombardment \citep{Spencer2020}.

The particle motion around an irregular-shaped minor body and their geophysical connection with the rotational Roche lobe is critical information providing clues on the surface evolution. For example, \citet{Mao2020} show that Arrokoth could lose mass from its surface. This behavior is expected because the escape speed across the surface of Arrokoth \citep{Amarante2020} is much lower than a typical impact speed among KBO's \citep{Greenstreet2019}. Besides that, the contact binary KBO Arrokoth has a unique slope pattern. For example, the direction of downslope motion for surface particles goes from the Arrokoth equator towards the poles of the lobes. This pattern suggests that some peculiar topographic regions on its surface are predisposed to accumulate or lose materials \citep{Amarante2020,Winter2020,Moura2020}. Therefore, the regions within the body's rotational Roche lobe limit may eventually trap particles from these materials \citep{Amarante2021,Scheeres2019,Scheeres2019b}. A study of particles' fate in the dynamical environment near Arrokoth's surface could then give insights about its surface topographic features for perception of the responsibility that pristine Kuiper Belt objects may have played in the process of the planetary formation.

For example, \citet{Rollin2020} show that the surface of Arrokoth contact binary absorbs particles if they have pericentric distance $\sim 2\times$ less than the distance between the centre of the Arrokoth's lobes. They used a dumbbell contact-binary shape for Arrokoth. In this work, assuming that the Arrokoth contact binary has a polyhedral shape, we study the mechanical environment's dynamics near its surface. We explored the orbital dynamics of massless particles around the surface of Arrokoth, considering its irregular binary geopotential, uniform density, and slow spin rate. Since there are four equilibrium points around Arrokoth's surface \citep{Amarante2020}, the dynamics of particles near Arrokoth's environment might be strongly connected to the topological structure of those equilibria. Note that the equilibrium points lie inside the circumbinary chaotic zone of Arrokoth \citep{Rollin2020}.

Using the most recent available three-dimensional (3-D) shape model of Arrokoth, we aim to investigate the dynamics of simulated particles around its surface, taking into account the stability of the regions close to the equilibrium points of the system. The goal is to answer some questions about the dynamics of this contact binary system. For example, since the simulated particles after some integration time could collide with the surface of each lobe: where are the selected regions for agglomeration of particles? How long do they keep their orbital motion around the surface of Arrokoth before impacting each lobe? How does the preferred impact sites are related to the initial inclinations of the simulated particles? How do the falls are associated dynamically with peculiar surface features of lobes, as MD impact crater? So, we performed samples of numerical experiments to explore the behavior of the fate of particles in the dynamical environment around the surface of the Arrokoth contact binary. 

This manuscript is structured as follows. In Section \ref{sec:math}, presented are the 3-D shape model used for the Arrokoth and the mathematical model adopted in our dynamic simulations. The mascons technique is used to find the irregular geopotential around the surface of the Arrokoth contact binary for the numerical integrations. We have also shown the solar radiation pressure equations for a comparative numerical experiment. At the end of this section, we reproduced the approximated location and topological stability of Arrokoth's equilibria for our chosen model. The following sections show the results of our sets of dynamic simulations. Three different types of initial conditions are considered to deal with the environment close to the surface of Arrokoth. They are presented in Section \ref{sec:sim}. In Section \ref{sec:sim:lod}, particles are initially distributed in local disks in the neighborhoods of the equilibrium points around Arrokoth. We also investigate synchronous orbits placing simulated particles at the approximated location of equilibrium points of the system. We also have tested a set of dynamic simulations that handle a ring of particles around Arrokoth that encompass all equilibria. This set of initial conditions are discussed in Section \ref{sec:sim:to}, where we also present the flux of particles in the environment around Arrokoth and the particle size thresholds. The last numerical experiment type is presented in Section \ref{sec:fall}: the impact sites over the surface of each lobe. In this type of sample, the simulated particles are initially located in a spherical cloud around the Arrokoth contact binary for a full range of inclinations. Finally, in Section \ref{sec:final} presented are our final remarks.

\section{Numerical Model}
\label{sec:math}
\begin{figure*}
  \centering
  \fbox{\includegraphics[width=18cm]{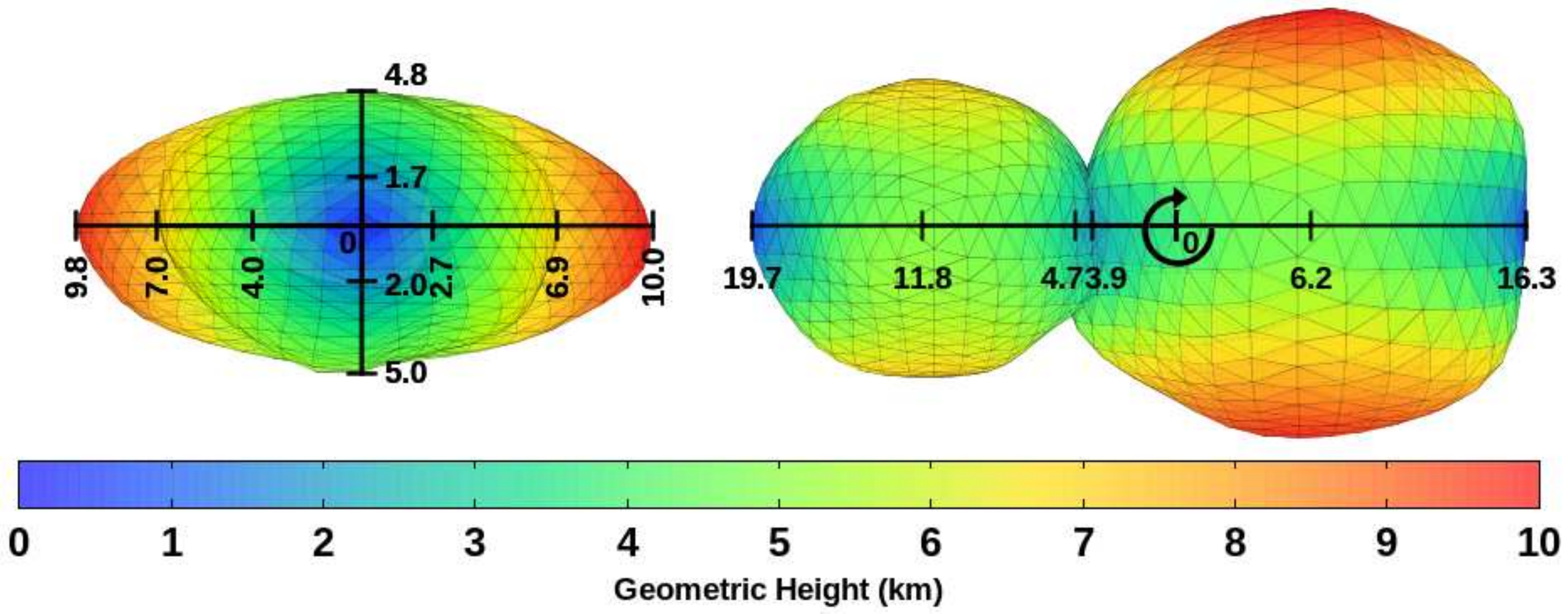}}
  \caption{(left-hand side) The dimensions of the lobes of Arrokoth contact binary. The bars denote the sizes of the `neck'; small and large lobes, measured from major axis $x$ (0), in km. (right-hand side) The 3-D polyhedral shape model of Arrokoth contact binary as seen by \textit{New Horizons}'s spacecraft at close encounter. The bars give the distances from the barycentre of Arrokoth contact binary (0) along the $x$-axis, in km. The angular velocity vector has the direction of the semiminor axis $-z$ according to the clockwise movement, passing through the body centre of mass (0). The geometric height is represented by the colour bar.}
  \label{fig:math_2}
\end{figure*}
\begin{figure}
  \centering
  \fbox{\includegraphics[width=8.44cm]{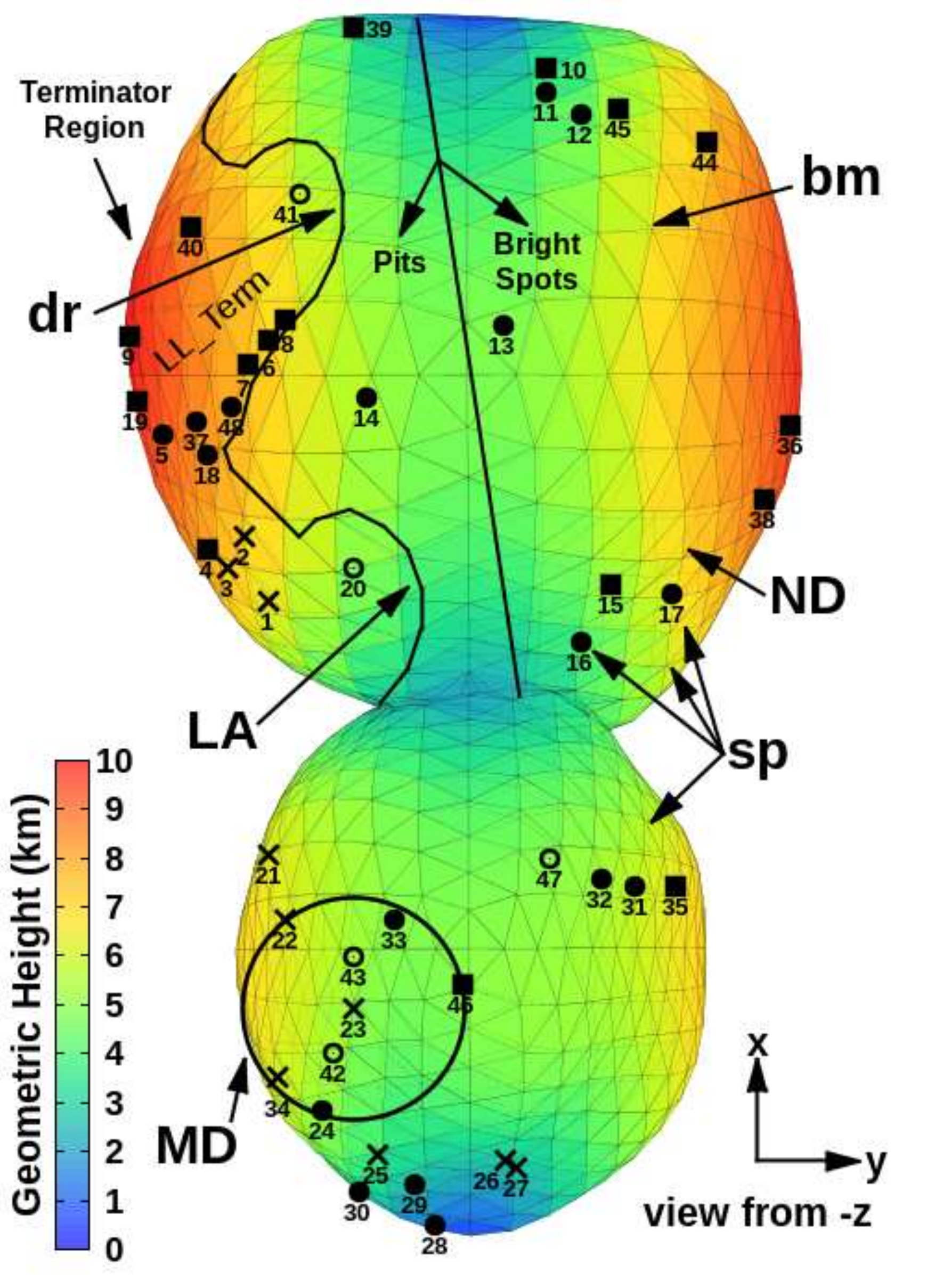}}
  \caption{Craters and Pits on Arrokoth. The colour bar indicates the geometric height. Label numbers refer to crater listings in Data S3 from \citet{Spencer2020}. Point type denotes confidence class: X-dots = high confidence, full circle-dots = medium confidence, square-dots = low confidence. Features indicated in circle-dots are considered to be highly unlikely to be of impact origin. The solid straight black line splits the large lobe into regions with different lighting conditions, a more obliquely-illuminated region with more visible depressions (Pits) and a more vertically-illuminated region with bright spots (Bright spots). The curved black line delineates the boundary of combined geologic units (LL\_Term), considered together for crater density determination \citep{Spencer2020}. Moreover, the black circle under the small lobe's surface represents the approximated location of the Maryland crater (MD). Abbreviations and informal feature nicknames are: MD = Maryland impact crater, LA = Louisiana, dr = depressed region, bm = bright material, ND = North Dakota, and sp = spots \citep{Grundy2020}.}
  \label{fig:math_2b}
\end{figure}

The numerical model adopted here considers the most important perturbation for particles around minor bodies far away from the Sun: the irregular geopotential generated by the bilobate shape model of Arrokoth. For the sake of comparison, we also did dynamic simulations for three different particle range sizes when the Solar Radiation Pressure (SRP) was considered. As we shall see further on, we are interested in the evolution and stability of simulated particles near the surface of Arrokoth. So, it is also essential to handle with a good representative shape model for its surface features, which has been developed by \citet{Stern2019,Spencer2020}. Thus, the orbital motion of particles around the Arrokoth contact binary can be explored accurately through this approach. The mathematical model used in this work is discussed in detail in the following subsections.

\subsection{Shape model}
\label{sec:math:shape}
A precise shape model of the Arrokoth contact binary is needed to study the dynamical environment's behavior near its surface. In this work, we used a low facet count version of the Arrokoth 3-D polyhedral shape model \citep{Spencer2020} performed for thermal models\footnote{The polyhedral data for Arrokoth is from the Science Supplementary Materials website: \url{https://science.sciencemag.org/content/suppl/2020/02/12/}\\\url{science.aay3705.DC1}.} \citep{Grundy2020}. This shape model is the most recent shape model of the surface of Arrokoth that was visible to \textit{New Horizons} near the closest approach, derived from the CA04 and CA06 LORRI images. For our purposes, we are only interested in the geometric vertices of \citet{Spencer2020}'s best-fitting shape model. So, in Fig. \ref{fig:math_2}, we rebuilt the Arrokoth's low facet shape model using $1,039$ geometric vertices and $2,999$ edges combined into $1,962$ triangular faces\footnote{The GNUPlot program \citep{Williams2011} is used to build the polyhedral mesh of this figure.}. The colour box code gives the geometric height, i.e., the distance between the centroid of each triangular face and the major axis $x$. In that way, the total volume of Arrokoth contact binary is 3,166\,km$^3$, equivalent to a sphere of radius 9.1\,km. This volume is 30.4\% larger than the previous estimate by \citet{Stern2019}, though consistent within the uncertainties. Comparing the current shape model with the previous one (e.g., see Fig. 1 from \citet{Amarante2020}), the small lobe has a more ellipsoidal shape (perspective views $\pm x$), and the large lobe is less flattened in the $z-$axis direction (perspective views $\pm y$). Note that the right-hand side of Fig. \ref{fig:math_2} is plotted in the same perspective seen by \textit{New Horizons}' spacecraft at close encounter time (perspective view $-z$). However, our Arrokoth model's rotation pole lies along the shortest $z$-axis (principal axis of the largest moment of inertia) according to the right-hand rule (counterclockwise motion). In Fig. \ref{fig:math_2b}, we also approximately locate each known feature on the surface of the Arrokoth \citep{Spencer2020, Grundy2020}, from \textit{New Horizons}' spacecraft perspective of view, using its 3-D polyhedral shape model.

\subsection{Binary geopotential}
\label{sec:math:geo}
One of the foremost techniques to model the geopotential of an irregular-shaped minor body is to consider its volume occupied with point masses (mascons) using a cubic grid spaced uniformly \citep{Geissler1996}. Thereby, the mascons model is particularly proper to handle minor bodies due to its capacity to model very irregular shapes with low-resolution \citep{Werner1997b,Scheeres1998,Park2010}.
Nevertheless, the mascon model for the computation of the gravitational field near the surface of an irregular-shaped minor body has accuracy problems \citep{Werner1997b}. In this work, we focus on the approximated locations where the simulated particles collided. Most of the particles that hit the surface of Arrokoth are placed initially close to the equilibrium point regions, where the use of the mascon model could be considered accurate enough to study its dynamical environment \citep{Amarante2021}.
This approach has a significant advantage due to the simple conceptual technique and low computational effort. With these assumptions the binary gravitational force potential $U_b(\textbf{r})$ can be computed by the sum of individual ones $U_1(\textbf{r})$ and $U_2(\textbf{r})$ of the large (1) and small (2) lobes, respectively:
\begin{align}
U_b(\textbf{r}) &= U_1(\textbf{r})+U_2(\textbf{r}) = \sum_{i=1}^{N_1}\frac{Gm_i}{|\textbf{r}-\textbf{r}_i|}+\sum_{j=1}^{N_2}\frac{Gm_j}{|\textbf{r}-\textbf{r}_j|},
\label{eq:math_1}
\end{align}
\noindent where $N_1=10,107$ and $N_2=5,297$ denote the amount of point of masses confined into the volume of the large and small lobes, respectively. In this approach, we considered an approximated $590$\,m spaced mascon cubic cells for each lobe following its irregular shape. Thus, the mascon model for Arrokoth contact binary is composed by $N_1+N_2 = 15,404$ point masses with equal mass of $m_i = m_j = 1.027582 \times 10^{11}$\,kg each one, where $m_i=M_1/N_1$ and $m_j=M_2/N_2$. $M_1$ and $M_2$ are the total body mass of the large and small lobes given in the Tab. \ref{tab:math_1}. $Gm_i$ and $Gm_j$ are the gravitational parameters of each point of mass, where gravitational constant has the value\footnote{CODATA - \url{http://physics.nist.gov/constants}} $G=6.67408 \times 10^{-20}$\, km$^3$\, kg$^{-1}$\, s$^{-2}$. $\textbf{r}$ represents the particle radius vector from Arrokoth barycentre in the body-fixed frame, whose the unit vectors are defined along with the principal moments of inertia. $\textbf{r}_i$ and $\textbf{r}_j$ indicate the point mass distance relative to centre mass of the Arrokoth contact binary. $|\textbf{r}-\textbf{r}_i|$ and $|\textbf{r}-\textbf{r}_j|$ represent the distances of a particle from mascons of each lobe.

From the mutual binary gravitational force potential given by Eq. \eqref{eq:math_1} we compute the binary mascon gravity attraction vector as follows: 
\begin{align}
\nabla U_b(\textbf{r}) &= -\sum_{i=1}^{N_1}\frac{Gm_i}{|\textbf{r}-\textbf{r}_i|^3}(\textbf{r}-\textbf{r}_i)-\sum_{j=1}^{N_2}\frac{Gm_j}{|\textbf{r}-\textbf{r}_j|^3}(\textbf{r}-\textbf{r}_j),
\label{eq:math_2}
\end{align}
\noindent where $\nabla$ represents the gradient operator.

In that way, the binary gravity gradient tensor can be computed as shown in Eq. \eqref{eq:math_3}:
\begin{align}
\nonumber \nabla\nabla U_b(\textbf{r}) & = \pdv{U_b(x_1,x_2,x_3)}{x_m}{x_n} =\\& \sum_{i=1}^{N_1}\frac{Gm_i}{r_i^3}\Bigg[\frac{3(x_m-x_{m_i})(x_n-x_{n_i})}{r_i^2}-\delta_{m,n}\Bigg]+\nonumber
\\&\sum_{j=1}^{N_2}\frac{Gm_j}{r_j^3}\Bigg[\frac{3(x_m-x_{m_j})(x_n-x_{n_j})}{r_j^2}-\delta_{m,n}\Bigg]\nonumber\\&\,\,\,\,\,\,\,\,\,\,\,\,\,\,\,\,\,\,\,\,\,\,\,\,\,\,\,\,\,\,\,\,\,\,\,\,\,\,\,\,\,\,\,\,\,\,\,\,\,\,\,\,\,\,\,\,\,\,\,\,\,\,\,\,\,\,\, (m,n = 1,2,3),
\label{eq:math_3}
\end{align}
\noindent where $(x_1,x_2,x_3)\equiv(x,y,z)$ and $\delta_{m,n}$ is the Kronecker delta.

Consequently, the Arrokoth's binary geopotential is given by Eq. (\ref{eq:math_4}) \citep{Amarante2020}:
\begin{align}
V_b(\textbf{r}) &= -\frac{1}{2}\omega^2(x^2+y^2)-U_1(\textbf{r})-U_2(\textbf{r}),
\label{eq:math_4}
\end{align}
\noindent where $\omega=2\pi/T$ is the spin rate of the Arrokoth contact binary with the rotation period $T$ given from Tab. \ref{tab:math_1}.

The previous Arrokoth contact binary features (Eqs \eqref{eq:math_1}-\eqref{eq:math_4}) are numerically computed using the \textit{Minor-Gravity package}\footnote{\url{https://github.com/a-amarante/minor-gravity}.} \citep{minor-gravity}.
\begin{table}
\centering
  \caption{Orbital and physical properties of the Arrokoth contact binary system.}
\begin{threeparttable}
 \label{tab:math_1}
 \begin{tabular}{ccc}
  \toprule
   Parameter & Value & Units \\
  \hline
  Semimajor axis\tnote{\textit{a}} & $6.617545\times 10^9$ & km \\
  Eccentricity\tnote{\textit{a}} & 0.037874 & -- \\
  Inclination\tnote{\textit{a}} & 2.4499 & degrees \\
  Arg. of perihelion\tnote{\textit{a}} & 183.7480 & degrees \\
  Long. of asc. node\tnote{\textit{a}} & 339.0471 & degrees \\
  Mean anomaly\tnote{\textit{a}} & 301.3045 & degrees \\
  Obliquity\tnote{\textit{b}} & 98.0 & degrees \\
  Rotation period ($T$)\tnote{\textit{b}} & 15.92 & hours \\
  Large lobe mass ($M_1$) & $1.038577\times 10^{15}$  & kg \\
  Small lobe mass ($M_2$) & $5.443102\times 10^{14}$  & kg \\
  \hline
 \end{tabular}
\footnotesize
    \begin{tablenotes}
        \item[\textit{a}] \cite{Porter2018}.
        \item[\textit{b}] \cite{Stern2019}.
    \end{tablenotes}
\end{threeparttable}
 \end{table}
\begin{table}
\centering
  \caption{The SRP parameters used in the mathematical model.}
 \label{tab:math_2}
 \scalebox{1.0}
{
 \begin{tabular}{cccc}
  \toprule
  Parameter & Value & Units & Comments \\
  \hline
  $Q_{pr}$ & 1 & -- & \text{efficiency factor} \\
  $S_0$ & $1.36\times 10^3$ & \text{kg/s$^3$} & \text{solar constant} \\
  $R_0$ & $1.495979\times 10^8$ & \text{km} & astronomical unit \\
  $c$ & $2.997922\times 10^5$ & \text{km/s} & speed of light \\
  $\rho$ & 0.5 & \text{g/cm$^{3}$} & bulk density \\
  $r_p$ & 0.01--10 & $\mu$\text{m} & particles radius \\
  \hline
 \end{tabular}}
 \end{table}

\subsection{Solar radiation pressure}
\label{sec:math:srp}
The radial SRP force due to Sun also affects the dynamics of the particles in the environment around irregular-shaped minor bodies. This force becomes weaker in the small bodies' environment far away from the Sun, like the case of the Arrokoth contact binary system. However, even in a distant region such as a trans-Neptunian object's system environment, the effects of the SRP have to be considered to better estimate the orbital evolution of dust particles \citep{Pires2013}.

A particle moving under the influence of the SRP acceleration has equation of motion as follows \citep{Burns1979,Mignard1984}:
\begin{eqnarray}
\textbf{a}_{SRP} & = & -{3\over 4}\frac{Q_{pr}S_0R_0^2}{c|\textbf{r}_{sun}|^3\rho r_p}\textbf{r}_{sun},
\label{eq:math_5}
\end{eqnarray}
\noindent where $|\textbf{r}_{sun}|$ is the distance from the Sun to the orbiting particle and $\textbf{r}_{sun}$ is the position vector of the particle relative to the Sun. The definition of quantities $\rho$, $r_p$, $S_0$, $R_0$, $c$, and $Q_{pr}$ are presented in Tab. \ref{tab:math_2}.

The particles in the numerical model are modeled as spheres. We also consider a Kleperian orbit for the Sun around Arrokoth, in the body-fixed frame, which orbital elements are given by Tab. \ref{tab:math_1}. 

\subsection{Equilibria location and topological stability}
\label{sec:math:eq}
\begin{figure}
  \centering
  \fbox{\includegraphics[width=8.44cm]{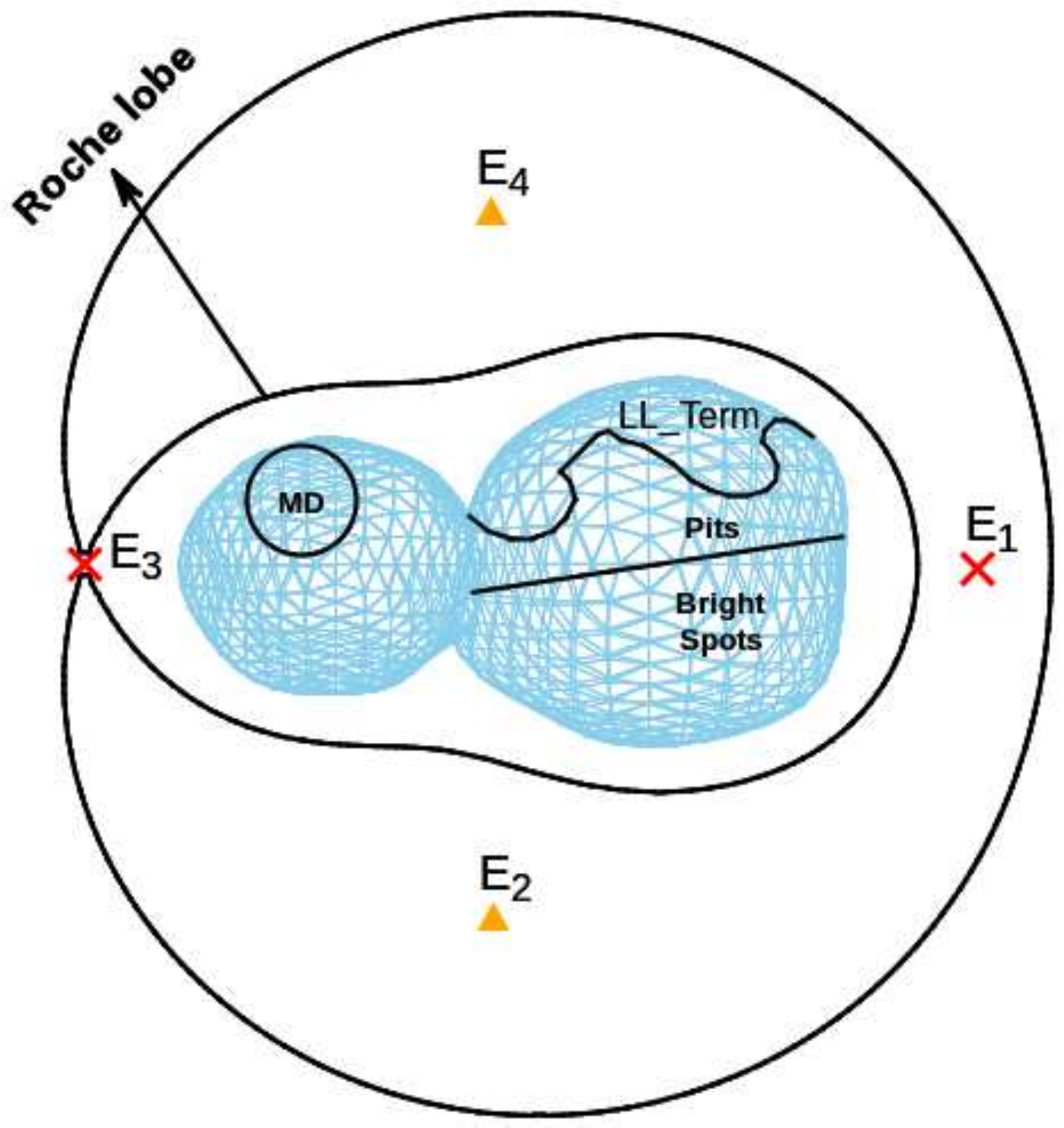}}
  \caption{Approximated location and topological stability of the equilibrium points, stationary points in the Arrokoth-fixed frame, projected in the equatorial plane ($-z$), for a constant density and a uniform rotation period. Red X-dots are topologically classified as hyperbolically unstable points (HU) and orange triangular-dots as complexly unstable points (CU). The rotational Roche lobe is drawn by the black contour line's outer region, which envelops the Arrokoth contact binary shape. (For interpretation of the references to colour in this figure legend, the reader is referred to the web version of this article.)}
  \label{fig:math_3}
\end{figure}

\begin{table*}
\centering
  \caption{Equilibrium points about KBO Arrokoth and their approximated location and topological stability. The definitions are the longitude $\lambda$, radial barycentric distance $r_{eq}$, the binary geopotential $V_b(x,y,z)$, the topological structure and the equilibrium point lifetime. HU = hyperbolically unstable points and CU = complexly unstable points. The equilibrium points were found through the mascons technique with \textit{Minor-Equilibria-NR package} \citep{minor-equilibria-nr} using an accuracy of $10^{-8}$. We also assumed a uniform density and spin rate for the Arrokoth contact binary.}
 \label{tab:math_3}
 \scalebox{1.0}
{
 \begin{tabular}{ccccccccc}
  \toprule
  Point & X (km) & Y (km) & Z (km) & $\lambda$ (deg) & $r_{eq}$ (km) & \specialcell{$V_b(x,y,z)$\\ \small(m$^2$/s$^{2}$)} & \specialcell{topological\\ stability} & \specialcell{equilibrium point\\ lifetime (h)} \\
  \hline
  E$_1$ & 23.4553 & -0.214880 & -0.0259389 & 359.475 & 23.4563 & -8.45575 & HU & 20 \\
  E$_2$ & -2.75162 & -19.0578 & 0.0299283 & 261.784 & 19.2554 & -7.33944 & CU & 35 \\
  E$_3$ & -24.7584 & 0.0277076 & -0.0616689 & 179.9359 & 24.7585 & -8.75538 & HU & 15 \\
  E$_4$ & -2.87830 & 19.0313 & -0.000448829 & 98.6002 & 19.2478 & -7.33984 & CU & 35 \\
  \hline
 \end{tabular}}
 \end{table*}


Considering the bilobate shape and rotation period of Arrokoth, \citet{Amarante2020} found four equilibrium points around it. The topological structure of them could be classified into hyperbolically and complexly unstable points. Therefore, the dynamical structure in the proximity of Arrokoth's surface is highly dependent on those equilibria and could affect particles' fate in the environment around the Arrokoth contact binary. In Table \ref{tab:math_3}, we show the topological stability and approximated location of the equilibrium points ($-\nabla  V_b(\textbf{r}) = \textbf{0}$) of the Arrokoth contact binary that were recomputed\footnote{\url{https://github.com/a-amarante/minor-equilibria-nr}.} for the most recent polyhedral shape model adopted in this work (Fig. \ref{fig:math_2}). The equilibria's approximated location will be used for the initial conditions of our dynamic simulations. Additionally, in Fig. \ref{fig:math_3}, presented are the equilibria approximated location and topological stability about Arrokoth. We also show the rotational Roche lobe limit of Arrokoth and approximated locations of some features of its surface defined in Fig. \ref{fig:math_2b}. It is important to mention that the rotational Roche lobe is attached to the minimum energy equilibrium point E$_3$. Depending on the adopted density, this particular equilibrium point moves towards equilibrium point E$_6$ to annihilate each other \citep{Amarante2020}. In that way, the rotational Roche lobe also could intersect the surface of the Arrokoth contact binary, likewise asteroid Bennu \citep{Scheeres2016}. 
Note that the equilibrium points in Fig. \ref{fig:math_3} and Tab. \ref{tab:math_3} are located in the clockwise direction because this figure was made in the same perspective of view ($-z$) of \textit{New Horizons} spacecraft's flyby.

\section{Dynamic Simulations}
\label{sec:sim}
The correspondence between the surface dynamics of an irregular-shaped minor body and its rotational Roche lobe (Fig. \ref{fig:math_3}) were shown by \citet{Scheeres2019,Scheeres2019b}. It can directly affect the evolutionary surface feature scenario of the Arrokoth contact binary. For example, the comprehension of the motion of particles near the Arrokoth's surface is critical to guide knowledge about their surface features' evolution. Also, its connection with the topological stability of the equilibrium points. The slope behavior on the surface of Arrokoth goes from equator to lobe poles \citep{Amarante2020}. Thus, the lobe poles have stable resting sites, as MD impact crater. On the other hand, the equatorial region of each lobe is unstable. It makes the surface stability of this body very peculiar. Considering the centrifugal acceleration, particles that leave from an unstable equatorial area could trap in orbits around Arrokoth inside the rotational Roche lobe.

In the current work, we are not concerned with the origin of the particles. Assuming hypothetical particles nearby Arrokoth, we studied their evolution and outcome. However, it is reasonable to expect some particle configurations around Arrokoth. Minor bodies were formed through collisions or ruptures because they were rapidly rotating. So, it is natural that, in the past, these bodies provided events of particles around them that may form satellites or even rings, as in the know cases of (10199) Chariklo \citep{Braga-Ribas2014} and (136108) Haumea \citep{Ortiz2017}. For example, the asteroid (101955) Bennu was seen ejecting particles from its surface by the OSIRIS-REx mission \citep{Lauretta2019}. \citet{Mao2020} show that Arrokoth could also lose mass from its surface. Therefore, the regions within the body's rotational Roche lobe limit may eventually trap particles from these materials in a planar (ring), inclined (cloud), or around an equilibrium configuration (local disks).

Then, three types of numerical experiments are considered to get information about the evolution and stability of particles in the dynamical environment close to the surface of the Arrokoth contact binary. They are:

(I) local disks,

(II) ring of particles,

(III) spherical cloud.

Types (I) and (II) are performed to understand how the stability of the equilibria near the equatorial region of Arrokoth governates the fate of particles around its surface. The discussions about them are made in the sections \ref{sec:sim:lod}--\ref{sec:sim:to}. Type (III) is run to explore the distribution of particles that impact the surface of each lobe. It is presented in Section \ref{sec:fall}. The numerical simulations have a full time of $1.14$\,yr (10,000\,h). Since the particles are initially distributed in the vicinity of Arrokoth's surface, most of them will impact the surface or escape from its environment in less than 1,000\,h. Our goals will be attended through numerical integrations using the \textit{Minor-Mercury package}\footnote{\url{https://github.com/a-amarante/minor-mercury}.} \citep{minor-mercury}. We used the N-body algorithm B\"{u}lirsch--St\"{o}er \citep{Stoer1980} in the numerical simulations. In the Appendix \ref{sec:minor-mercury} our collisional and escape criteria are presented. The following sections give details about each type of simulation. 

\section{The Local Disks}
\label{sec:sim:lod}
\begin{figure*}
  \centering
  \includegraphics[width=8.44cm]{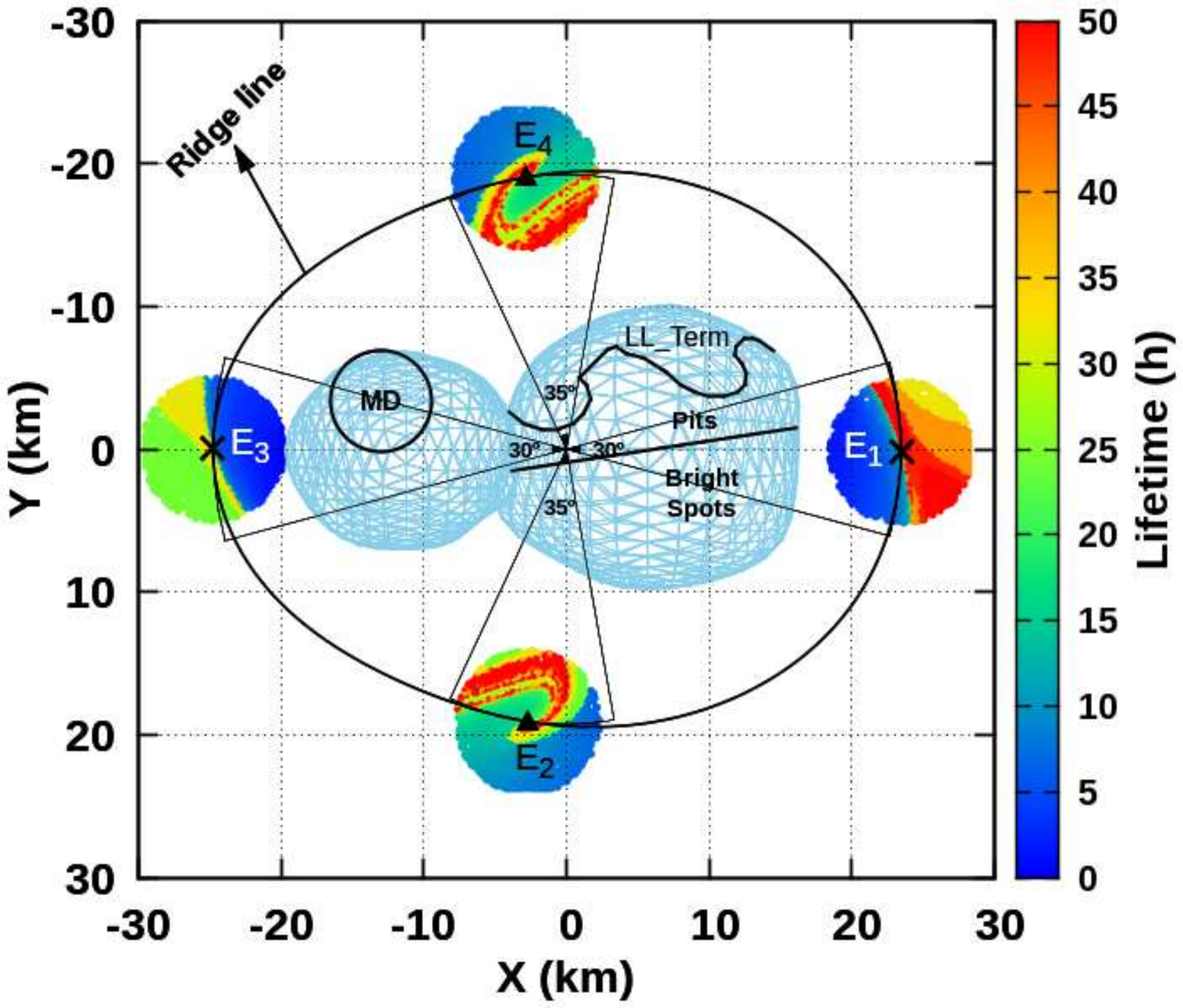}
  \includegraphics[width=8.44cm]{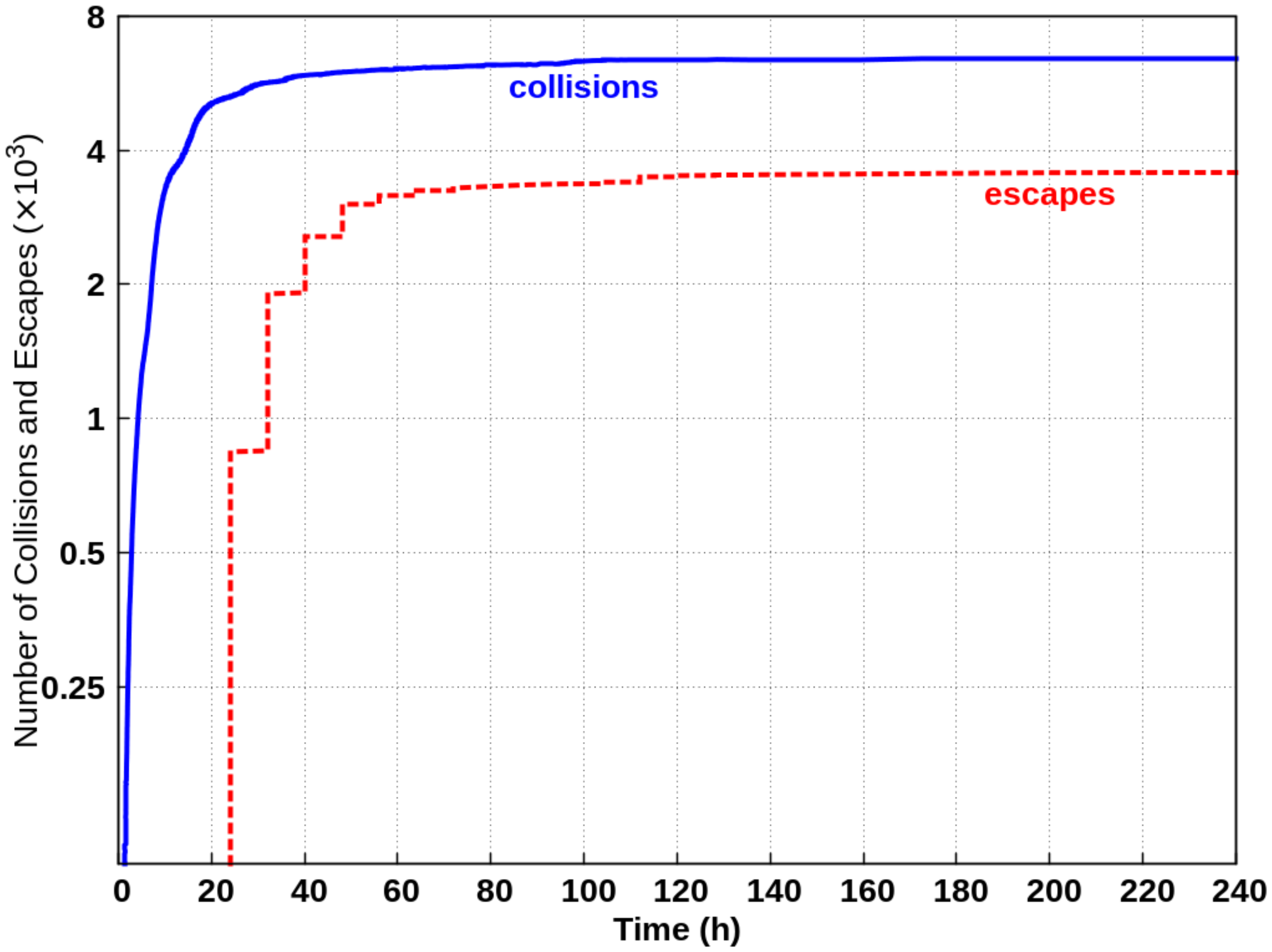}
  \includegraphics[width=8.44cm]{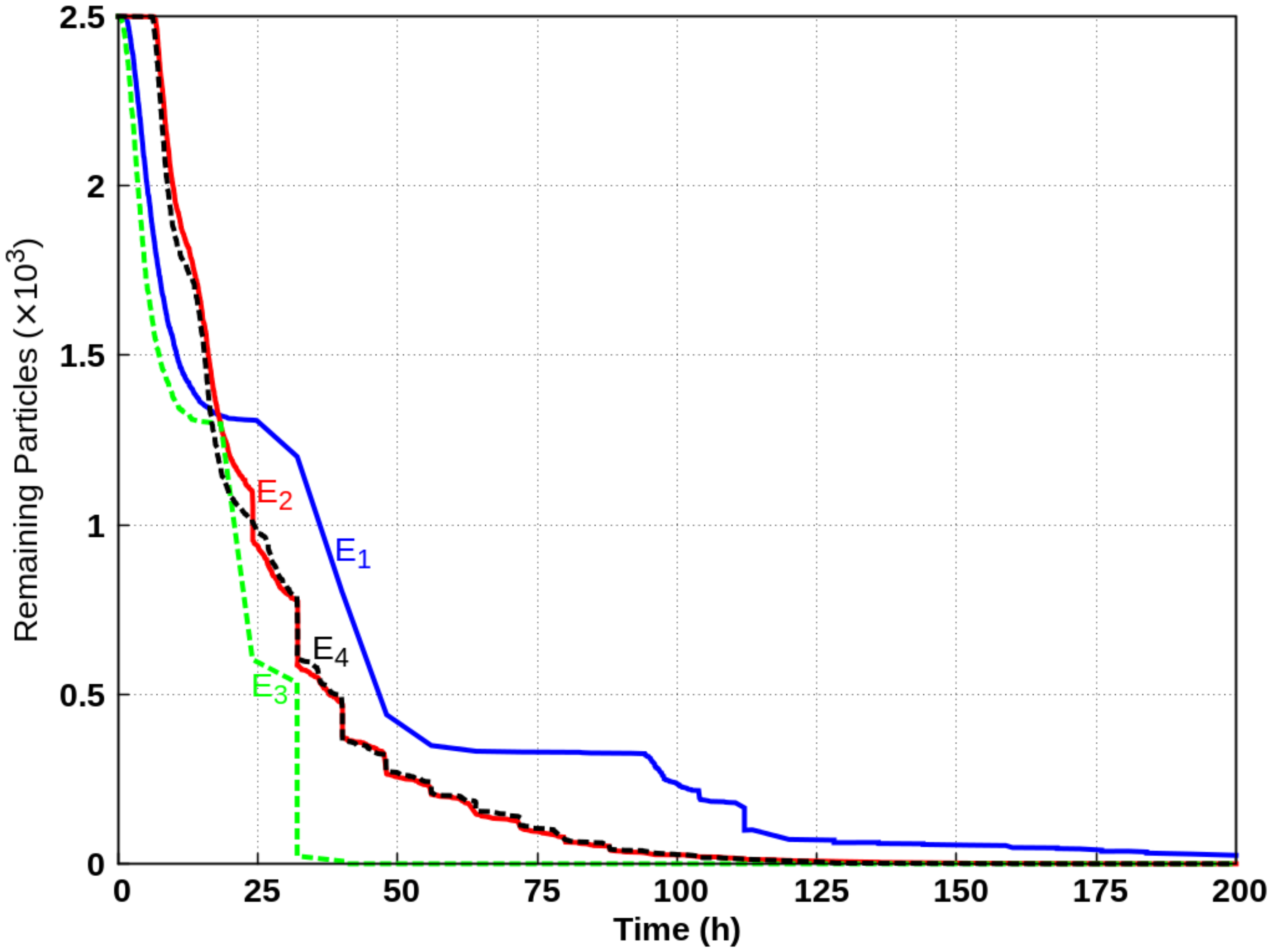}
  \caption{Numerical results for the simulated particles with initial conditions of type (I). (top-left side) Lifetime map of the initial location of each simulated particle that surrounds one of the four equilibria. The particle's lifetime is represented by the colour bar. (top-right side) The blue line indicates the number of collisions over time. The dashed red line denotes the number of escapes. (bottom side) The evolution of the number of the remaining particles in the local disks around equilibrium points E$_1$ (blue), E$_2$ (red), E$_3$ (dashed green), and E$_4$ (dashed black).}
  \label{fig:lod_1}
\end{figure*}

In this section, we explore how the topological stability of the equilibria could affect the fate of particles initially distributed in local disks in the neighborhoods of the equilibrium points around the Arrokoth contact binary. The local disks are filled evenly by 2,500 particles around each equilibrium point (E$_1$, E$_2$, E$_3$, and E$_4$). Each local disk has a radius equal to 5\,km centered at its corresponding equilibria location. The local disks are made with a large radius to cover a wide region of phase space around the equilibrium points. The radial barycentric distance of each equilibrium point is used to calculate the initial orbital speed of the corresponding sample of particles as a value proportional to it. In the Tab. \ref{tab:math_3} presented are the radial barycentric distance $r_{eq}$ of each equilibrium point. It was computed using Eq. \eqref{eq:to_1}, where was used $r_{eq}$ instead $r_{av}$. The particle radius vector is perpendicular to the particle velocity vector. We considered that the sample of particles was initially placed at the inertial frame.

The numerical results of type (I) numerical integrations presented are in Fig. \ref{fig:lod_1}. The top-left side of Fig. \ref{fig:lod_1} presents the lifetime map (denoted by the colour bar code) of the initial location of each simulated particle that surrounds one of the four equilibria. We do not consider the SRP perturbation in this type of numerical experiment. The black curve that passes through equilibria represents the ridge line of the Arrokoth contact binary. The ridge line consists of a closed smooth curve around an irregular-shaped minor body, where the radial and vertical components of the centripetal and gravitational forces are balanced. Beyond the ridge line, the effective acceleration pulls outwards; within the ridge line, the effective acceleration pulls inwards, towards the minor body. When the bulk density is decreased, the ridge line comes closer to the body surface \citep{Tardivel2014}.

The results show that the particles initially located in the neighborhoods of the surface of Arrokoth's lobes, such as surrounding hyperbolically unstable points E$_1$ and E$_3$, are removed very fast from the system in a couple of hours. However, there are some significant differences in the lifetime map due to the dynamic behavior of the ridge line. The particles within the ridge line distributed around local disks E$_1$ and E$_3$ (dark-blue dots) are more likely to be pruned from Arrokoth's environment than those in the neighborhoods of local disks E$_2$ and E$_4$ (from green to red dots). Meanwhile, the particles located initially inside the ridge line survive longer (red dots) for these two local disks. Otherwise, particles initially situated outside the ridge line in the local disks E$_2$ and E$_4$ are fast removed from Arrokoth's vicinity. Apart from that, particles in the local disk E$_3$ survive for intermediate times (there are no red dots). This local disk has three distinguishable regions for particles's lifetime: blue ($ 18$\,h), yellow ($ 24$\,h) and green ($ 32$\,h). On the other hand, there are four remarkable regions for lifetime maps consering particles initially placed in the local disk E$_1$: blue ($ 25$\,h), yellow ($ 32$\,h), orange ($ 40$\,h), and red ($ > 50$\,h).

The evolution of the number of collisions and escapes for the four local disk samples are shown in the top-right side of Fig. \ref{fig:lod_1}. Since the local disks are built with a large radius covering a wide area of phase space around the equilibrium points, most of the particles are pruned from the environment around Arrokoth due to impacts with its surface. They cover $\sim 64\%$ of the total removed particles from the system, while the $\sim36\%$ account for escapes. They start to escape significantly only after $24$\,h. 

The evolution of the number of remaining particles for the local disks E$_1$ (blue), E$_2$ (red), E$_3$ (dashed green), and E$_4$ (dashed black) are given in the bottom side of Fig. \ref{fig:lod_1}. The results indicate that more than $50\%$ of their sample of particles will hit the surface of the Arrokoth contact binary just in a couple of tens of hours ($<25$\,h). This feature is independent of the local disk. Local disks E$_1$ and E$_3$ also had their particles being escaped from the system until they reach half of their initial number of particles ($<32$\,h). Apart from that, particles that leave from local disk E$_1$ are pruned slower from the vicinity of the Arrokoth. This behavior continues before 18\,h, when particles from the local disks E$_2$ and E$_4$ start to be removed more slowly from the environment around Arrokoth. After this time, the local disk E$_1$ remains with 52\% of its initial amount of particles.

At the beginning of the integration, the sample of particles from local disk E$_3$ decreases fastest until 17\,h, followed by the local disk E$_1$ up to 25\,h. They remain with approximately half of their particles most due to collisions with the surface of Arrokoth. After 17\,h the remaining number of particles from local disk E$_3$ reached 24\% only in the next 7\,h, followed by the local disk E$_1$ that remains with the higher number of the total amount of particles (32\%) up to 40\,h (both due to escapes).

After 17\,h, local disk E$_4$ shrinks slightly faster than local disk E$_3$ up to 18\,h. Finally, after 40\,h, the sample of particles from local disk E$_3$ reaches zero. At this moment, particles from both local disks E$_2$ and E$_4$ are pruned quickly from the vicinity of the Arrokoth contact binary. Note that they are the slowest to be removed at the beginning of the integration. The sample of particles of these two local disks keeps almost the same until the end of their integrations ($< 190$\,h). For long-time integrations, few particles remain in the Arrokoth's neighborhood for the local disk E$_1$. The sample of particles initially situated in the local disk around equilibrium point E$_1$ stays in the numerical integration for more than 200\,h. They cover only 1\% of its initial number of simulated particles.

\begin{figure*}
  \centering
  \includegraphics[width=8.44cm]{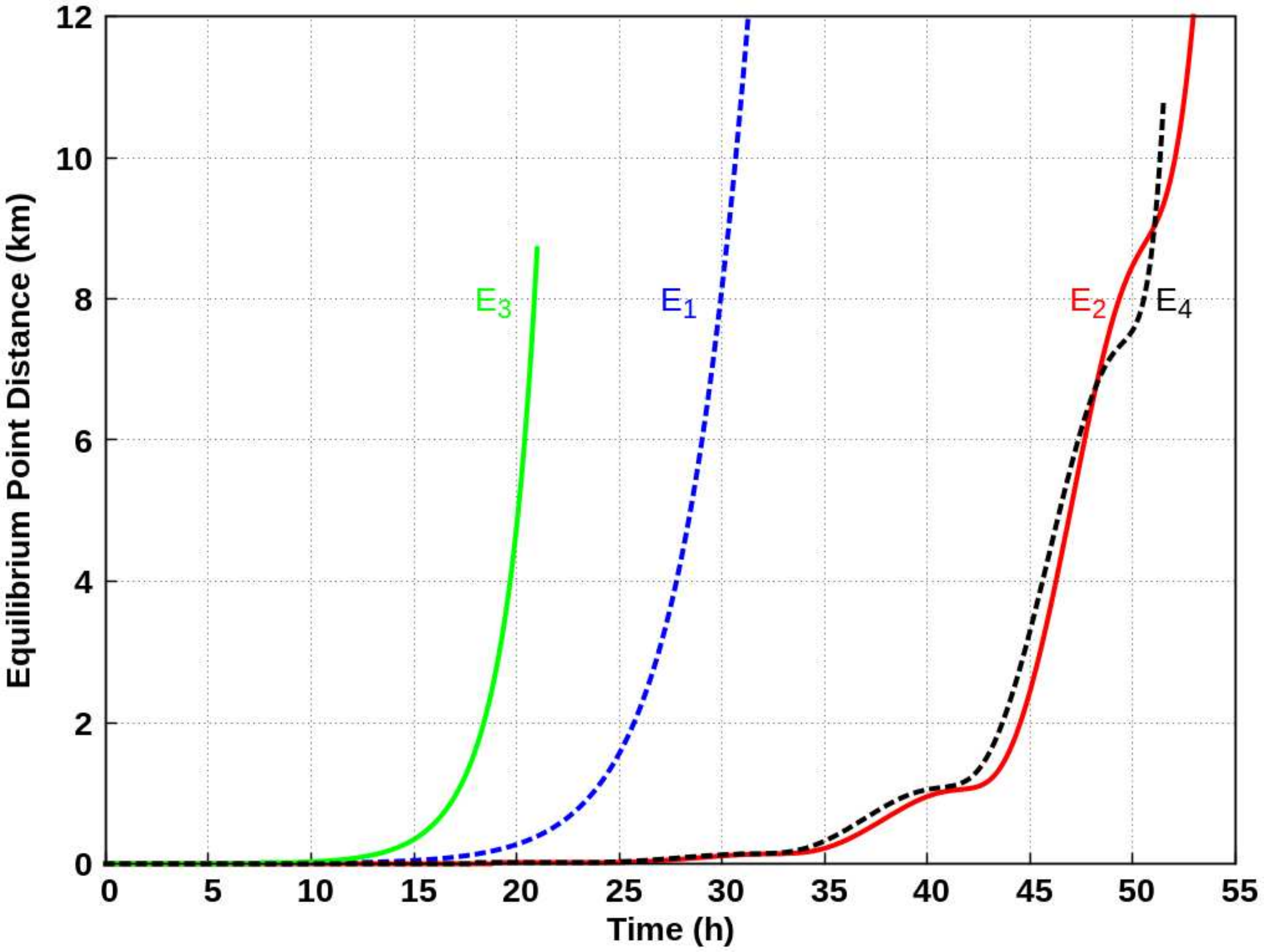}
  \includegraphics[width=8.44cm]{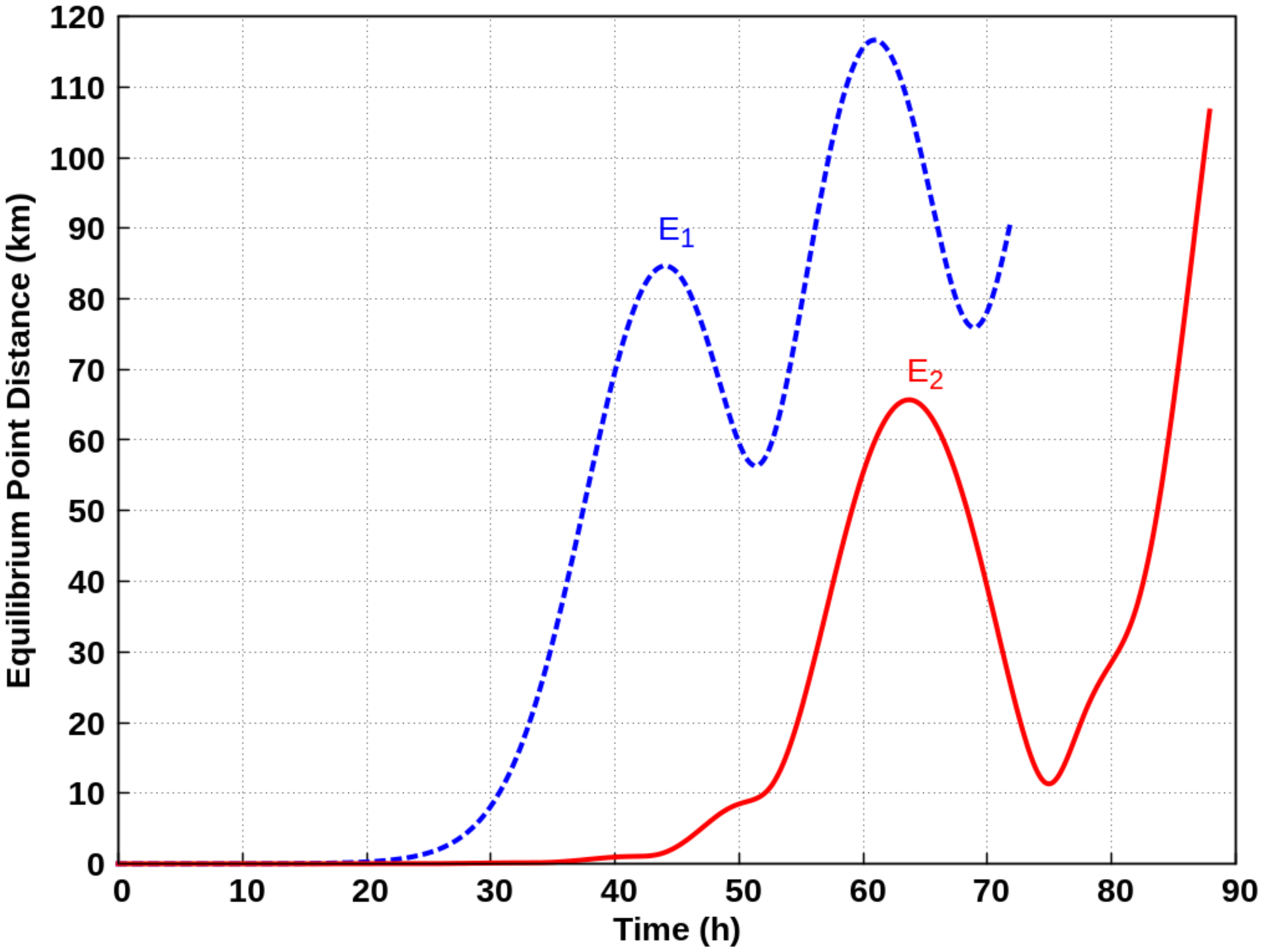}
  \caption{The evolution of particle distances from equilibrium points around Arrokoth contact binary. The particles are initially situated at the approximated location of each equilibrium point around Arrokoth in the body-fixed frame. (left-hand side) Distance from equilibrium point for particles with synchronous orbits initially located at the approximated location of equilibrium points E$_1$ (dashed blue), E$_2$ (red), E$_3$ (green), and E$_4$ (dashed black). (right-hand side) Distance from equilibrium point for particles with synchronous orbits initially situated at the equilibrium point locations E$_1$ (dashed blue) and E$_2$ (red).}
  \label{fig:lo_1}
\end{figure*}

We performed an animated movie, available online, that shows the behavior of the fate of simulated particles in the local disks for the initial conditions of type (I) (Movie 1). Note in this animation that the particles in the local disks E$_3$ and E$_1$ have a preference to reach the MD crater longitudinal region of the small lobe and Bright spots edge area of the large lobe, respectively. Furthermore, local disks E$_4$ and E$_2$ have their particles reached the LL\_Term area of the large lobe and the area diametrically opposite to it, of the small lobe, respectively.

\subsection{The synchronous orbits}
\label{sec:lo}

In this section, we placed a single particle at the computed location (Tab. \ref{tab:math_3}) of each equilibrium point to analyze the evolution and stability of the synchronous orbits around the Arrokoth contact binary. The particles are initially placed in the body-fixed frame. Thus, we compute the evolution of the synchronous particle orbits that depart from each equilibrium point until they hit the surfaces of the small and large lobes or escape from the neighborhood of Arrokoth.
Note that the location of an equilibrium point around an irregular-shaped minor body could be only computed numerically within an imprecision error. In this work, they are numerically computed considering the accuracy of $10^{-8}$ using the Newton-Raphson method \citep{minor-equilibria-nr}. Also, the evolution of the synchronous particle orbits is integrated with B\"{u}lirsch--St\"{o}er algorithm with a tolerance of $10^{-12}$ \citep{minor-mercury}. So, the evolution of the synchronous particle orbits that depart from each equilibrium point will be tested to give a general dynamical picture of the numerical equilibrium stability very close to the exact location of an equilibrium point.
The particles that are initially situated at the approximated location of an equilibrium point stay for an amount of time surrounding it before leaving its vicinity. This feature is due to the stability of the equilibria, that depending on its topological structure.
We define this amount of time numerically as the equilibrium point lifetime, and it is shown in Tab. \ref{tab:math_3}.
To better explore the behavior of synchronous orbits in the environment around Arrokoth, the particle distance from its corresponding equilibrium point is computed every timestep.

In Fig. \ref{fig:lo_1} presented are the results for the synchronous orbits. On the left-hand side of Fig. \ref{fig:lo_1}, we can see that the particle initially at equilibrium point location E$_3$ (green line) collides firstly with the small lobe surface, at $21.1$\,h. It departs from the neighborhood of equilibrium point E$_3$ at $\sim 15$\,h (equilibrium point lifetime). Note that this particle has the minimum value ($-8.76$\,m$^2$/s$^2$) between all binary geopotentials V$_b$ (Tab. \ref{tab:math_3}) and it also has a trajectory towards the equatorial region of the MD impact crater site on the small lobe before impacting near it. As we shall see further on, this impact area is following the regions of high values for the flux of particles in the environment around Arrokoth just before the particles impact its surface, as shown in the right-hand side of Fig. \ref{fig:to_3} (see animated Movie 3).

After 20\,h, the particle at equilibrium point E$_1$ (dashed blue line) is the second to depart from its equilibrium point. This particle evolves in the Arrokoth contact binary environment until it escapes from the system at 72\,h (right-hand side of Fig. \ref{fig:lo_1}).

After 35\,h, the particle from the equilibrium point E$_4$ (dashed black line) surrounds the equilibrium point until to collide with the Arrokoth's surface at 51.6\,h. This particle impacts the LL\_Term equatorial area of the large lobe.

Finally, the particle located at equilibrium point E$_2$ (red line) has an equilibrium lifetime of $\sim 35$\,h. This particle equilibrium lifetime is approximately the same as the particle which departs from equilibrium point E$_4$. Note that the binary geopotential V$_b$ from these equilibrium points (Tab. \ref{tab:math_3}) has approximately the same value ($-7.34$\,m$^2$/s$^2$). These particles keep approximately the same equilibrium point distance up to 51.6\,h (left-hand side of Fig. \ref{fig:lo_1}). The particle that departs from equilibrium point E$_2$ is also the last to remain in the simulation until to escape from the system at 88\,h (right-hand side of Fig. \ref{fig:lo_1}).

We also made an animated movie, available online, that shows the behavior of the fate of synchronous orbits in the environment around the Arrokoth contact binary (Movie 2).

\section{The Ring of Particles}
\label{sec:sim:to}
\begin{figure*}
  \centering
  \includegraphics[width=8.44cm]{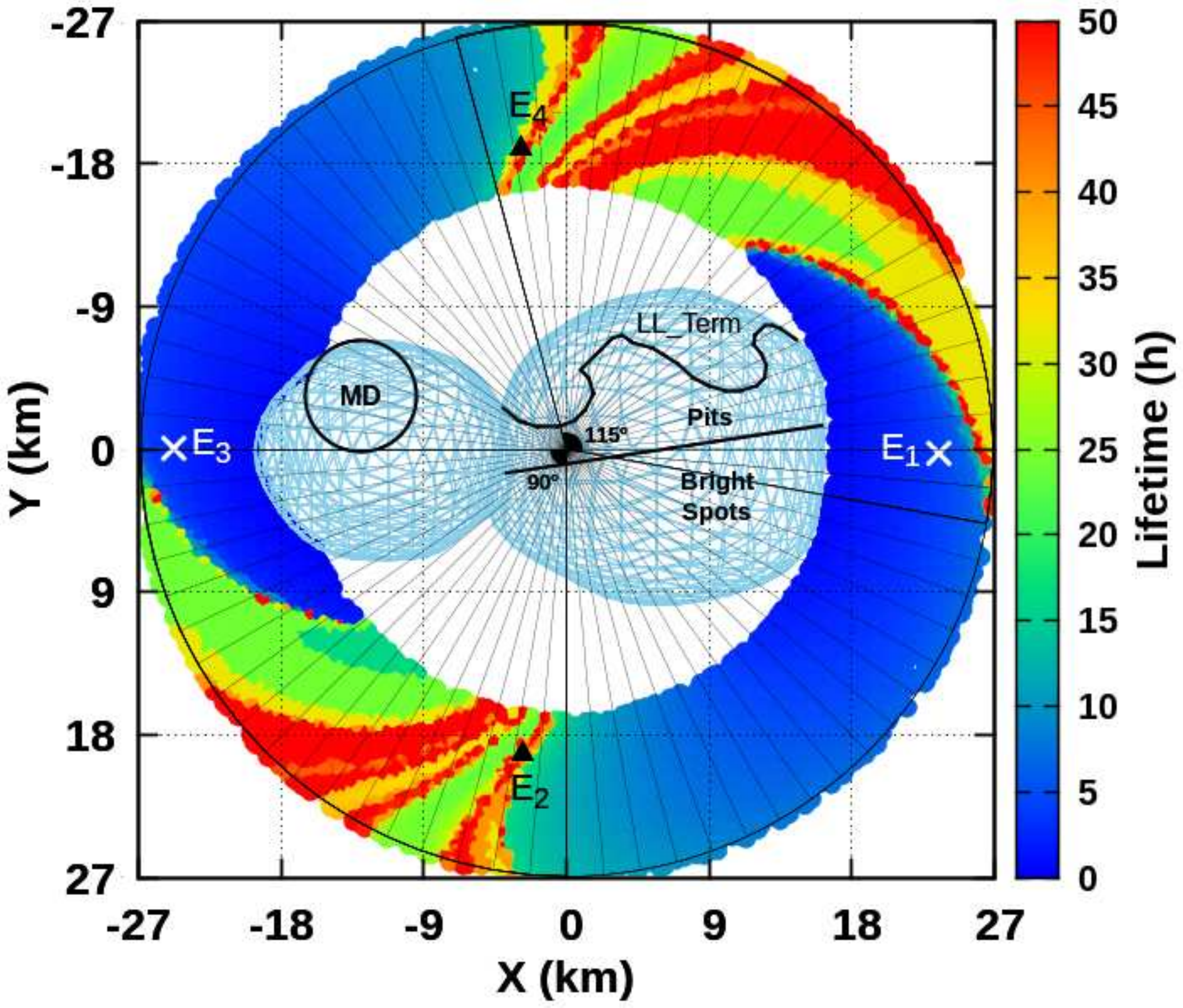}
  \includegraphics[width=8.44cm]{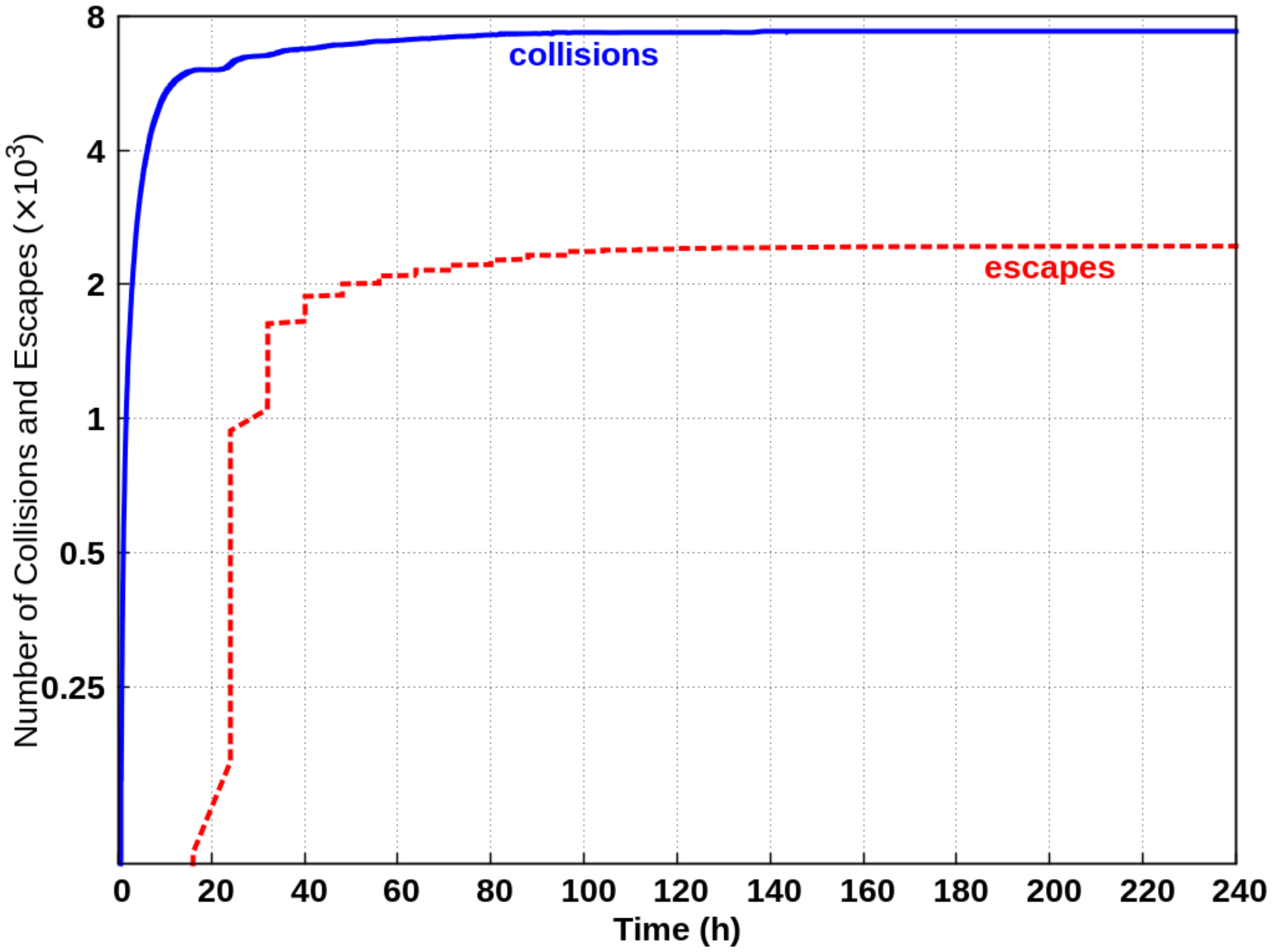}
  \includegraphics[width=8.44cm]{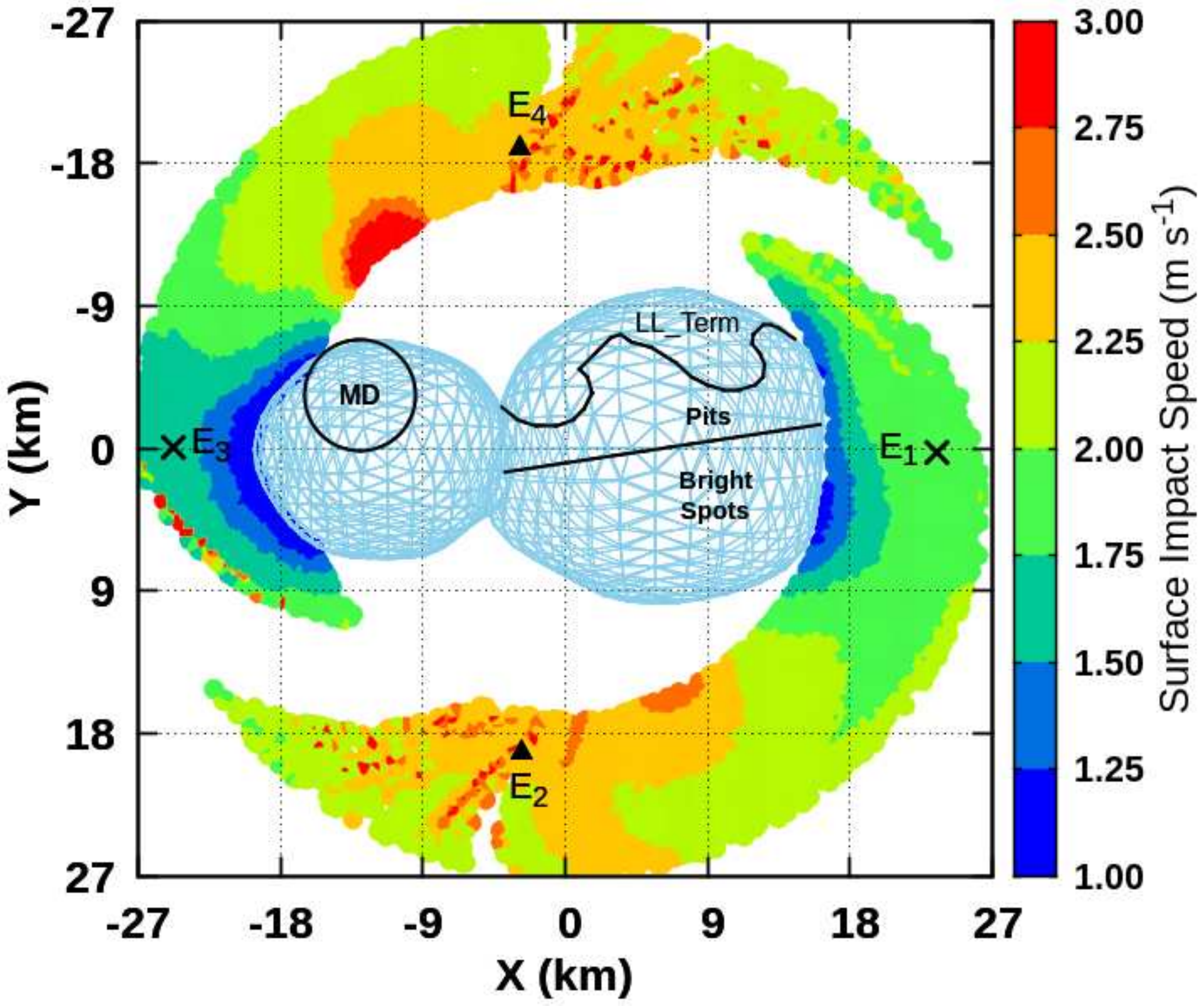}
  \includegraphics[width=8.44cm]{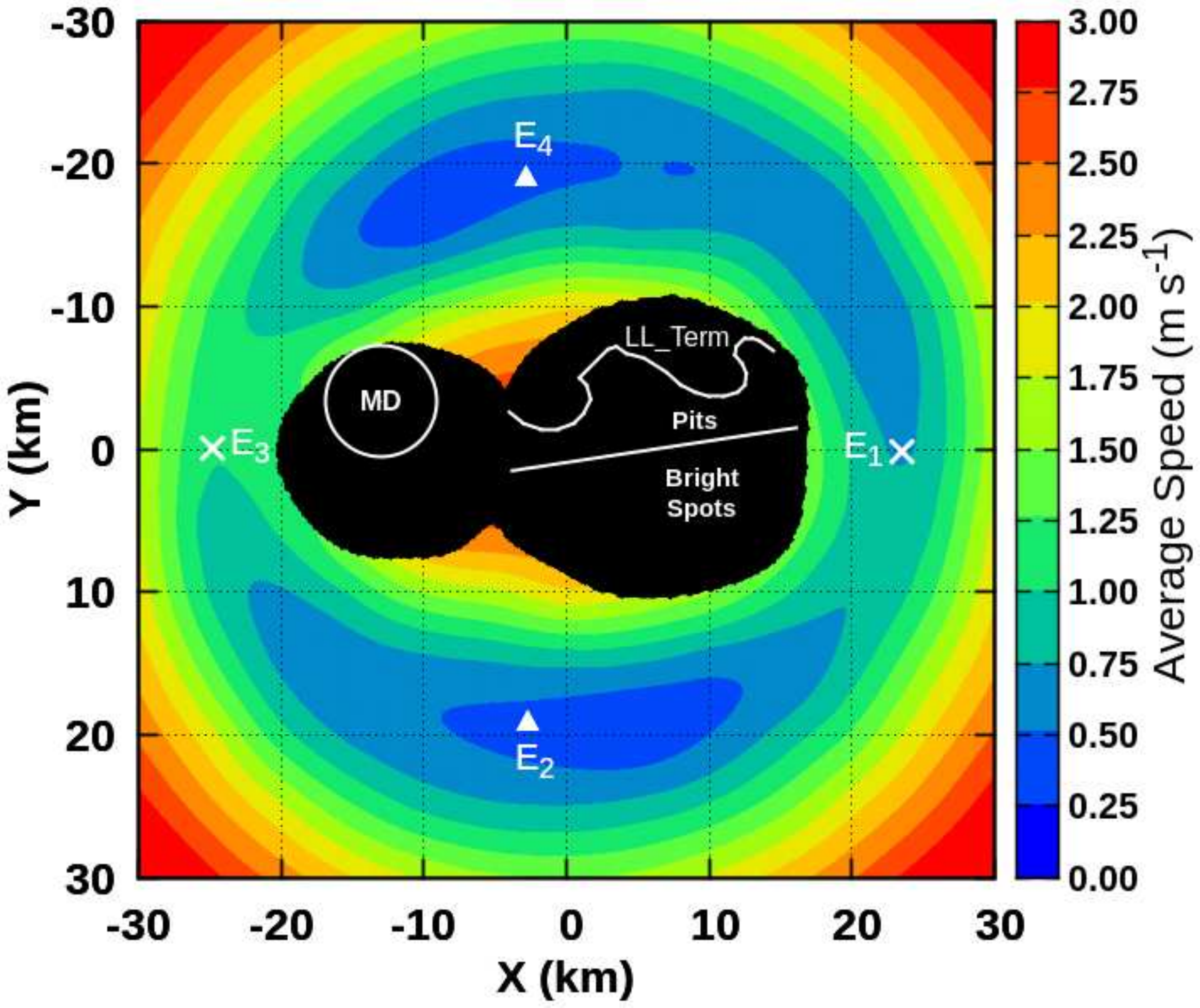}
  \caption{Numerical results for the simulated particles with initial conditions of type (II). (top-left side) Lifetime map of the initial location of each simulated particle. The particle's lifetime is represented by the colour bar. (top-right side) The blue line indicates the number of collisions over time. The dashed red line denotes the number of escapes. (bottom-left side) Surface impact speed of the particle's initial condition. The colour box gives the surface impact speed of each particle. (bottom-right side) Regions around Arrokoth contact binary for particle's average speed over the entire integration. All plots are made in the body-fixed frame.}
  \label{fig:to_1}
\end{figure*}
\begin{figure}
  \centering
  \includegraphics[width=8.44cm]{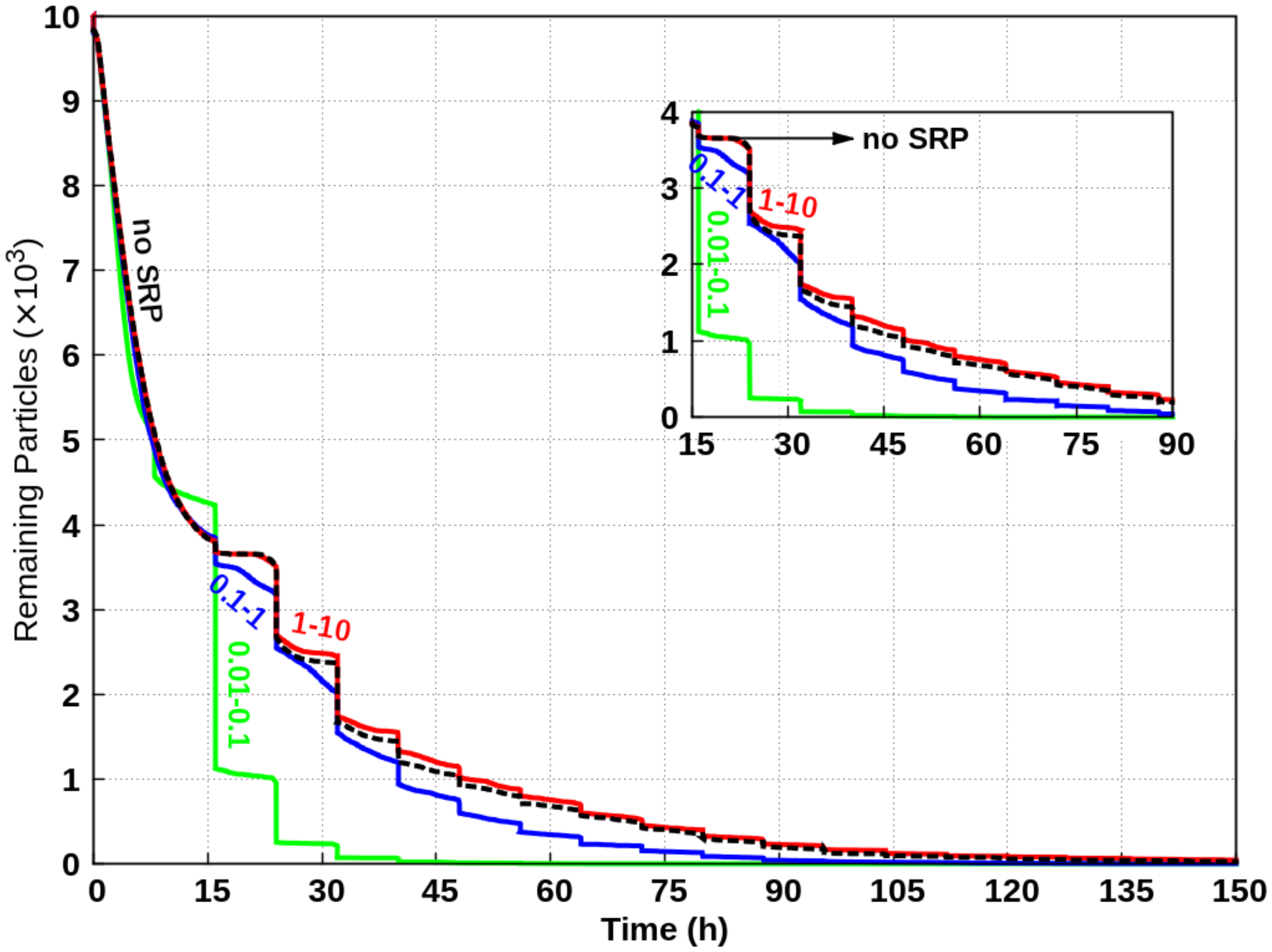}
  \caption{All sizes of the simulated particles of type (II) integrations with the evolution of the number of remaining particles. The colour line shows the interval of the particle size:
  Green, 0.01--0.1\,$\mu$m;
  Blue, 0.1--1\,$\mu$m;
  Red, 1--10\,$\mu$m.
  The black dashed line indicates the run of particles without the SRP perturbation.}
  \label{fig:to_2}
\end{figure}
\begin{figure}
  \centering
  \includegraphics[width=8.44cm]{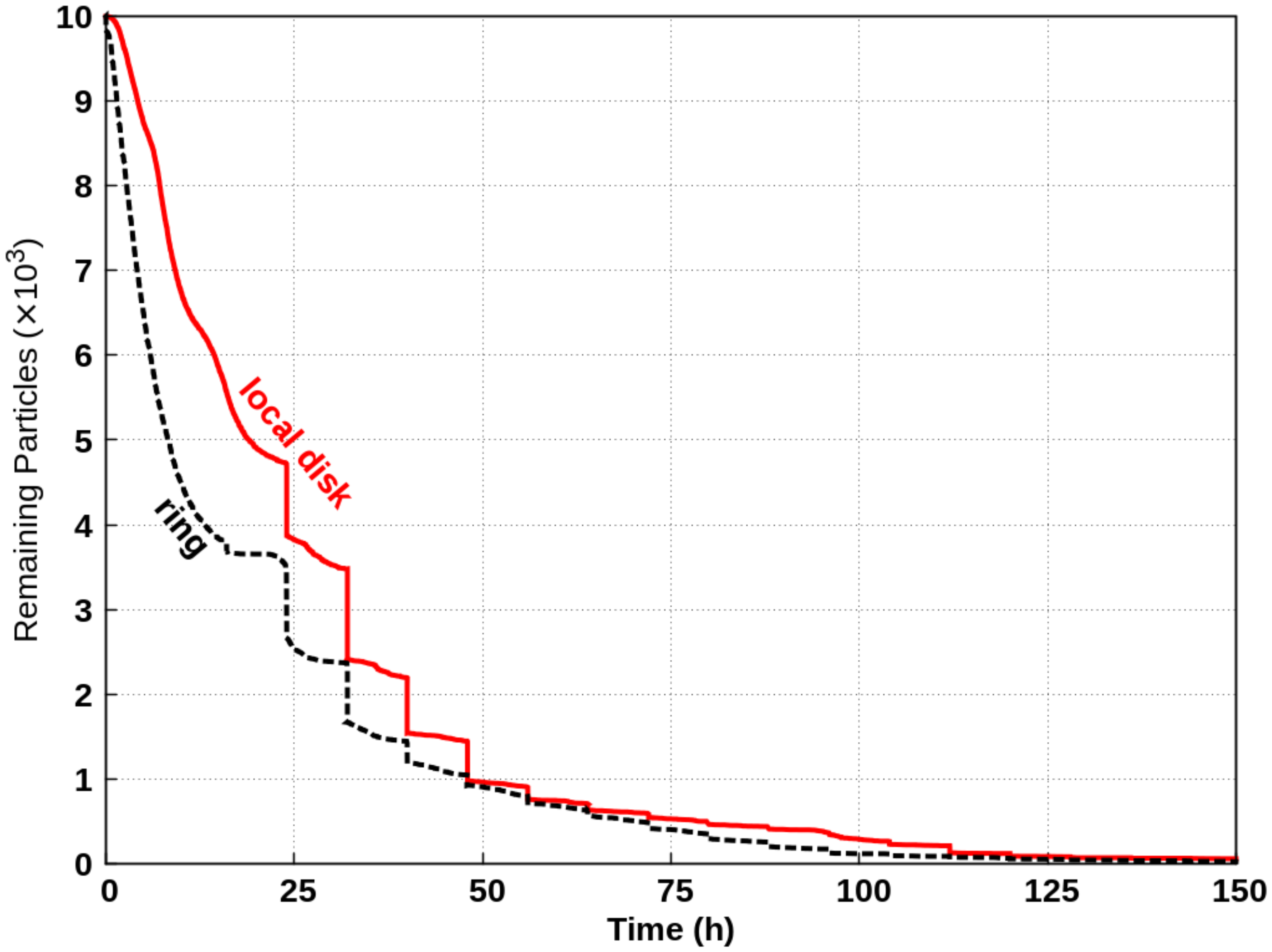}
  \caption{Comparing the evolution of the number of remaining particles for the sum of particles in the local disks (red) and the ring (dashed black) simulations, types (I) and (II), respectively.}
  \label{fig:lod_1b}
\end{figure}
In this type of initial condition, we distributed a uniform ring of particles around the Arrokoth contact binary using polar coordinates. The ring was made to involve the equilibrium points in the numerical simulations. The ring width lies between 17\,km and 27\,km from Arrokoth centre mass. The lower limit of 17\,km is chosen to fill Arrokoth's rotational Roche lobe and the regions close to odd equilibrium points E$_1$ and E$_3$ of particles. In comparison, the upper one (27\,km) considers the dominant region by the even equilibrium points E$_2$ and E$_4$. We distributed 10,000 particles uniformly around the surface of Arrokoth within the limits of the ring. The particles have initially circular orbits. They are set in the body-fixed frame using the \textit{HNM-Ring package}\footnote{\url{https://github.com/a-amarante/hnm-ring}} \citep{hnm-ring}. The mean anomaly and longitude of the ascending node were distributed randomly between $0^\circ$ and $360^\circ$. They also had an initial orbital speed value proportional to the average radial barycentric distance from equilibria (E$_1$, E$_2$, E$_3$, and E$_4$) at the inertial frame. We present the radial barycentric distance $r_{eq}$ of each equilibrium point in Tab. \ref{tab:math_3}. In that way, the average radial barycentric distance from equilibrium points has a value of 21.6831\,km. The particle radius vector is perpendicular to the particle orbital velocity vector. In that way, the initial orbital speed for a simulated particle is given by: 

\begin{eqnarray}
v_{os} & = & \omega \sqrt{\frac{r_{av}^3}{r}},
\label{eq:to_1}
\end{eqnarray}
\noindent where $r_{av}$ and $r$ are, respectively, the average radial barycentric distance and the particle radius from Arrokoth contact binary's centre mass.

We made an animated movie, available online, that shows the behavior of the fate of simulated particles around Arrokoth for the initial conditions of type (II) (Movie 3).
On the top-left side of Fig. \ref{fig:to_1}, the map of the lifetime of the particles is shown for initial conditions of type (II). This plot represents the initial location of particles, where the colour corresponds to their lifetimes. The ring is spread around the Arrokoth contact binary in a few hours (see animated Movie 3 available online). This behavior is expected from the dynamics since the ring is inside the circumbinary chaotic zone of Arrokoth \citep{Rollin2020}. Here, we focus on the lifetime of the particles affected by the stability close to the equilibrium points. In addition, we investigated the areas across the surfaces of individual lobes, which are reached by the simulated particles. Most particles are very fast pruned from the environment around Arrokoth contact binary in $\leq 50$\,h, as seen in the dark-blue region of this figure. There are two noticeable longitudinal regions around equilibria, where the particles remain in the system long times ($> 50$\,h). One of these regions is situated surrounding equilibrium points E$_1$--E$_4$, having an angular size of $115^\circ$. The other one is a $90^\circ$ longitudinal region near the equilibrium points E$_3$--E$_2$. The particles in the proximity of the MD crater site and along its longitudinal equatorial region (E$_4$--E$_3$) are pruned very fast from the system. Diametrically opposite to the longitudinal region between equilibrium points E$_2$--E$_1$ this feature can also be noted, where Bright spots and equatorial areas of Pits are located. Otherwise, most particles that remain long times before impacting with Arrokoth's surface or escaping from the system have initial locations within longitudinal equatorial areas near the LL\_Term region and diametrically opposite to it. It suggests longitudinal symmetric stability about a ring close to the equatorial area around the Arrokoth contact binary.
\begin{figure*}
  \centering
  \includegraphics[width=8.44cm]{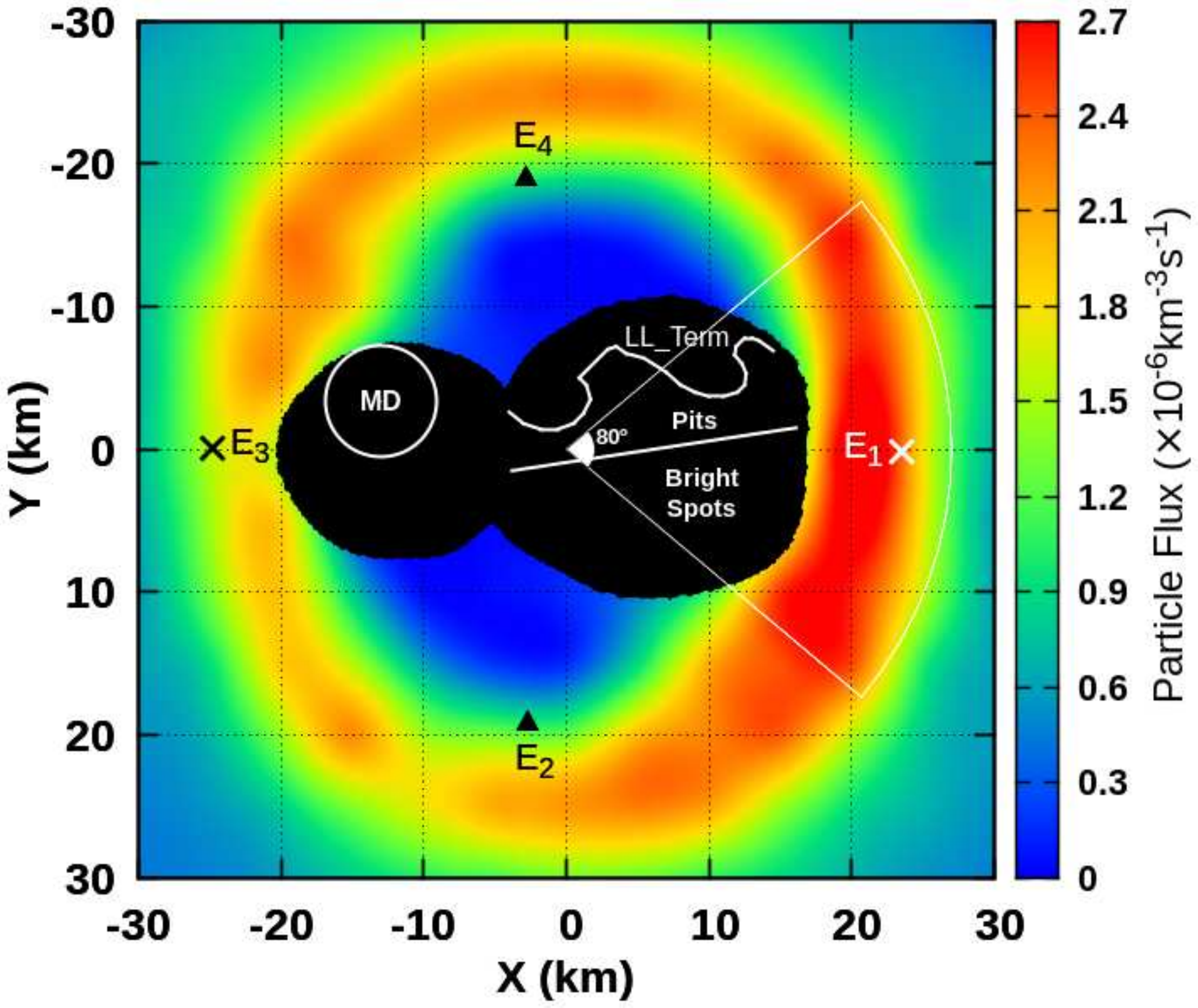}
  \includegraphics[width=8.44cm]{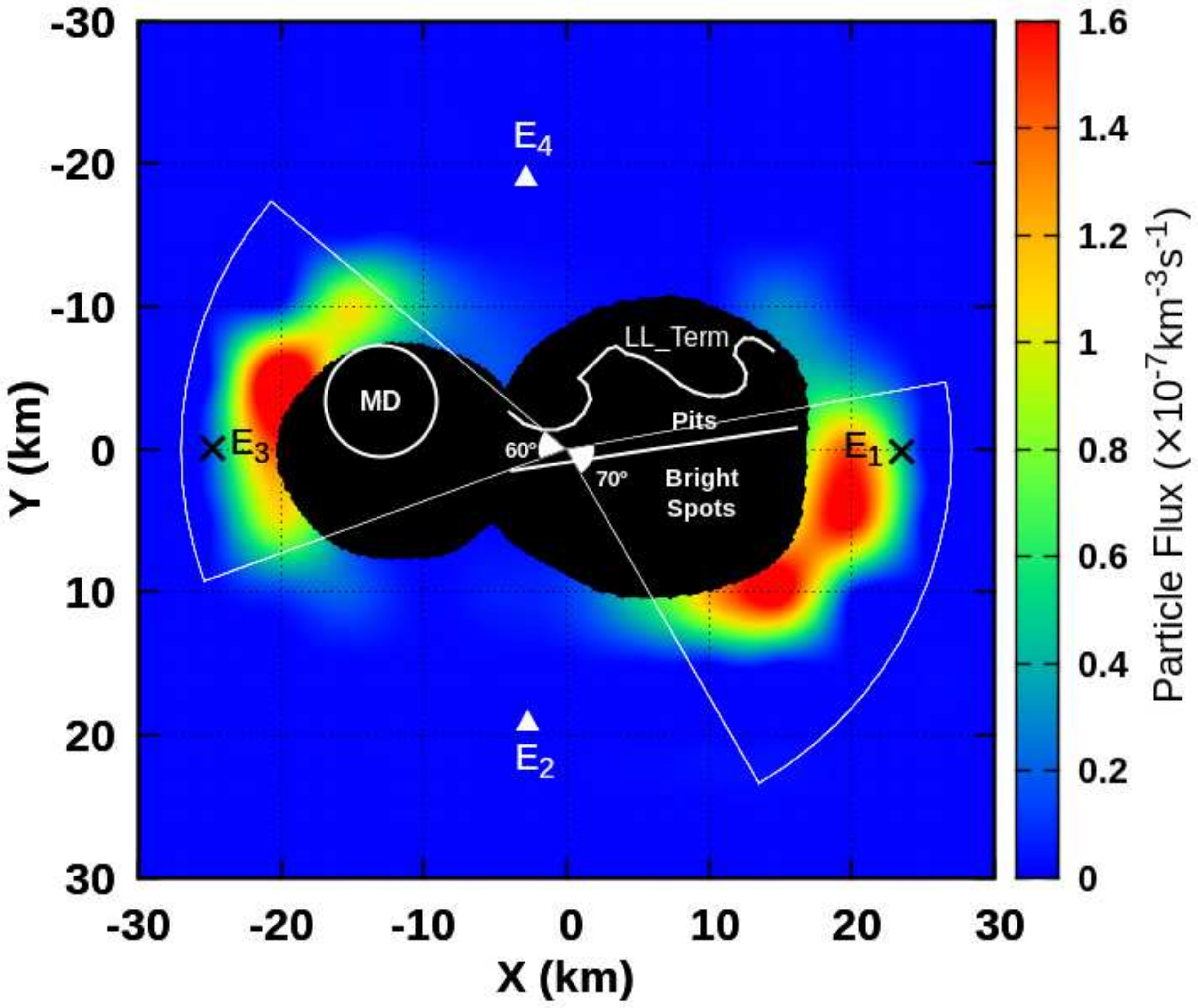}
  \caption{Numerical results for the simulated particles with initial conditions of type (II). (left-hand side) The map of the flux of particles in the environment of Arrokoth contact binary averaged over the entire integration. The particle flux is denoted by the colour bar code. (right-hand side) The particle flux map around Arrokoth averaged over the last 30\,min, just before the particles collided with the surface of Arrokoth. These are initial conditions for type (II) numerical integrations using the body-fixed frame.}
  \label{fig:to_3}
\end{figure*}

The number of collisions and escapes over time are shown in the top-right side of Fig. \ref{fig:to_1}. This plot is made considering the sample of particles for type (I) initial conditions. The ring around Arrokoth loses half of its initial number of particles in less than $ 9$\,h. The particles start very close to the Arrokoth's environment. Thus, most of them are pruned from the system due to impacts on its surface. From the total number of removed particles, $\sim 3/4$ is due to collisions with Arrokoth's surface. The particles start to escape from the system only after $16$\,h. They cover $\sim 1/4$ of the removed particles.

The surface impact speed around Arrokoth is plotted on the bottom-left side of Fig. \ref{fig:to_1}. Again, this figure is made at the body-fixed frame for the simulated particles of type (I) numerical integrations. The gaps are regions of the initial conditions where the particles are escaped from the system. The particles fall to the Arrokoth contact binary surface with impact speeds $\leq 3$\,m/s, agreeing with the contact velocity of the Arrokoth evolution from a binary planetesimal into contact by Kozai-Lidov oscillations and nebular drag \citep{Lyra2020}. Most particles impact the surface of Arrokoth with speeds in the range between 1.5 and 2.5\,m/s (from dark-green to yellow). Particles with speeds of 1 to 1.5\,m/s reach the equatorial area near the MD crater site (small lobe). In the LL\_Term equatorial region (large lobe) and diametrically opposite to it, there are particles ($\sim 25$\%) that escaped from the system (gaps). This behavior suggests that speeds $>3$\,m/s are more likely to force particles to escape from the environment around Arrokoth. These escaped particles have typically lifetimes between 20--40\,h (top-left and top-right sides of Fig. \ref{fig:to_1}). We also plot the particles' average speed over the entire simulation in the bottom-right side of Fig. \ref{fig:to_1}. The particles' average speed behavior in the vicinity of the surface of Arrokoth shows that most of the particles stay in the environment of Arrokoth with average speeds less than 3\,m/s, considering the body-fixed frame. Also, the particles' average speeds have most values between 1 and 2.25\,m/s before impacting to Arrokoth contact binary's surface.

\subsection{The size of particles}
\label{sec:sim:srp}

The simulated particles in the type (II) initial conditions are also affected by the SRP. We considered numerical experiments of particles with sizes from 0.01 up to 10\,$\mu$m distributed into three different intervals: $0.01-0.1$\,$\mu$m, $0.1-1$\,$\mu$m, and $1-10$\,$\mu$m. The initial number of particles of each size interval was 10,000, considering a uniform distribution. We adopted two criteria for removing a particle from the system. The particle is pruned if it collides with the surface of Arrokoth or it keeps its motion far away from the Arrokoth's environment (see Appendix \ref{sec:minor-mercury}).

Figure \ref{fig:to_2} shows the evolution of the number of the remaining particles affected by SRP. The line colours denote the sample of particles of the same size. These results indicate that more than half of the initial number of particles is removed from Arrokoth's environment in less than $8$\,h. This behavior occurs independently of the particle size. $\sim 88\%$ of particles are pruned from the proximity of Arrokoth very fast up to $16$\,h. They cover particles with sizes $\leq 0.1$\,$\mu$m. Most of them escaped from Arrokoth's neighborhood due to the significant effect of the SRP. It is important to mention that electromagnetic forces dominate particles with sizes $\ll 0.01$\,$\mu$m around minor bodies, and gravity becomes a perturbation \citep{Mignard1982,Horanyi1996}. We do not include electromagnetic forces in this work. In contrast, particles with sizes between 1 and 10\,$\mu$m are not significantly perturbed by SRP (black dashed line). This fact suggests that particles superior to a few microns are neatly dominated by the irregular gravity field of the Arrokoth contact binary. For comparison purposes, a sample of 1\,$\mu$m-sized particles in the environment of the Pluto-Charon binary system is also completely removed from the system due to SRP in a very short time-scale \citep{Pires2013}.


Figure \ref{fig:lod_1b} shows the difference between the total amount of remaining particles in the simulations type (I) and (II), the local disk integrations and the ring, respectively. There are more escapes for the initial conditions of type (I) than in the case where the particles are distributed in a ring around Arrokoth without SRP perturbation (type (II)). The ring shrinks faster than local disks at the beginning. After $\sim 50$\,h, both simulations remain with almost the same amount of particles up to 150\,h, where is slightly higher the number of remaining particles in the local disks.

\subsection{The flux of particles}
\label{sec:sim:flux}
The flux of particles is defined as the number of particles that pass through a unit volume averaged over time. Figure \ref{fig:to_3} shows the map of the flux of particles in the environment around Arrokoth for integration type (II). The flux of particles is represented by the code of colours in the boxes. We count the number of particles passed through a given box of the space around the Arrokoth contact binary overtime. The same particle can pass through the box several times during the integration. For example, on the left-hand side of Fig. \ref{fig:to_3}, the reddish regions correspond to places where a higher number of particles stayed for a long period.

In contrast, in the bluish regions, few particles stay along with the simulation. This behavior indicates the dynamic routes in the environment around Arrokoth for the motion of particles during the simulation. Depending on the particle orbital motion, it could keep or avoid preferred regions near the equilibrium points following the particle flux. It shows that the flux of particles has an accumulation around the equilibrium point E$_1$ (HU) and has a longitudinal range of $\sim 80^\circ$ that reaches the equatorial edges of the large lobe features (LL\_Term, Pits, and Bright spots).

Meanwhile, on the right-hand side of Fig. \ref{fig:to_3}, we count only the number of particles in the last 30\,min of the integration, just before they hit with Arrokoth's surface. The results show that most of the particles impact the surface of Arrokoth through longitudinal regions, where are situated equilibrium points E$_1$ and E$_3$. Note that the flux of particles has the higher values close to the MD crater site of the small lobe and the Bright spots area of the large one. This plot also indicates the approximated equatorial angular range for the accumulation of particles around each lobe, which is $70^\circ$ for the large lobe and $60^\circ$ for the small one.

This result agrees with the lifetime map of the particles around Arrokoth, which was built in Fig. \ref{fig:to_1}. This behavior shows that most of the particles initially distributed in a ring within the rotational Roche lobe of Arrokoth will have a fate over its surface through the longitudinal region of the MD crater site, located in the small lobe; or close to the region diametrically opposite to it. Conversely, a few of them will impact its surface in the longitudinal neighborhood of the LL\_Term area of the large lobe.
The animated Movie 3, available online, shows the behavior of the flux of particles during the simulation. Note that the dynamic routes of the particles pass through the equatorial region in the vicinity of the MD crater site. The particle flux also reaches the equatorial area diametrically opposite to it, i.e., the boundary of the Bright spots area. Besides that, most of the particles during the simulation pass through equilibrium point E$_1$ (HU) following the particle flux of Fig. \ref{fig:to_3}. Most of the particles reach the small lobe surface through equilibrium point E$_3$ (HU) and the surface of the large one through equilibrium point E$_1$. Also, there are particles that librate around the complexly unstable points E$_2$ and E$_4$.

\section{Distribution of Impacts}
\label{sec:fall}

In this section, simulated particles are initially distributed in a spherical cloud around the Arrokoth contact binary (type of initial conditions (III)). Our goal is to explore the distribution of impacts over the surface of each lobe. The SRP perturbation is not included in this numerical experiment since we are focused on particles that will hit Arrokoth's surface (larger than a few microns, see Fig. \ref{fig:to_2}).

\subsection{The spherical cloud}
\label{sec:sph}
The simulated particles are initially distributed uniformly in the environment around the Arrokoth contact binary, considering a spherical cloud-centered at Arrokoth's centroid.
The spherical cloud of particles has radial dimensions from 17\,km up to 47\,km, $\sim 2.6\times$ the distance between the centres of the lobes.
The spherical cloud is considered as a sample of 20,000 particles placed randomly around Arrokoth with circular orbits. The initial orbital inclination of the particles lies in the interval between $0^\circ$ and $180^\circ$. We investigated sample of particles with prograde (inclination $< 90^\circ$) and retrograde (inclination $> 90^\circ$) orbits. In addition, areal number density maps are performed to understand how the distribution of impacts across the surface of Arrokoth is related to some peculiar surface features. Beyond that, the particles in the spherical cloud have initial speeds $<2.5$\,m/s below their guaranteed return speeds \citep{Amarante2020}. So, this is sufficient for our purposes and consistent with the energy levels of Arrokoth's rotational Roche lobe thresholds.
\begin{figure*}
  \centering
  \fbox{\includegraphics[width=\linewidth]{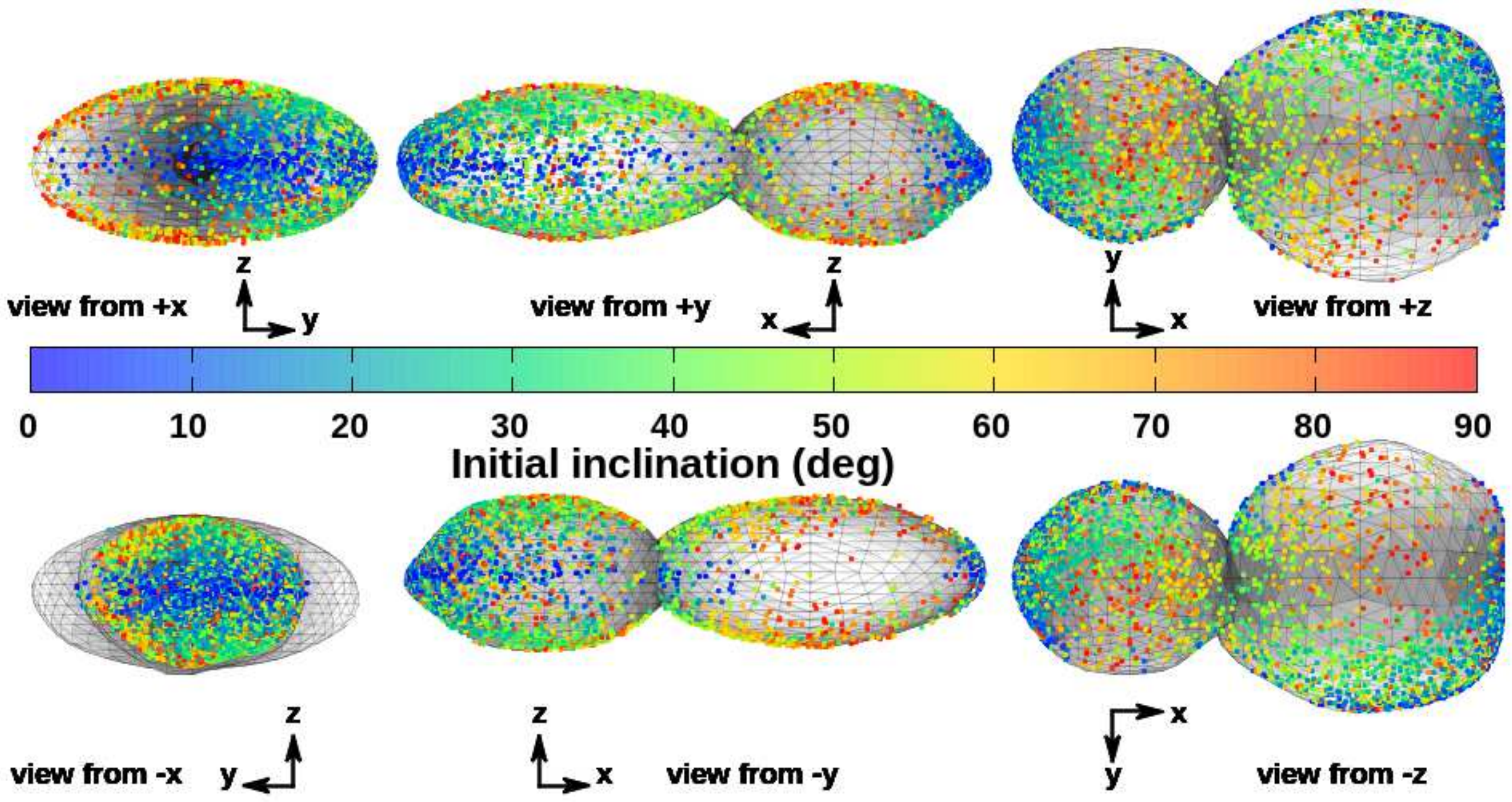}}
  \caption{Map of distribution of impacts over Arrokoth's surface. Prograde impact sites are represented by the dots, where their colours indicate the initial inclination of the particle orbit in the inertial frame. The geometric height of the surface of Arrokoth is given by the colour gray code. These are numerical results for the simulated particles with initial conditions of type (III).}
  \label{fig:fall_1}
\end{figure*}
\begin{figure*}
  \centering
  \fbox{\includegraphics[width=\linewidth]{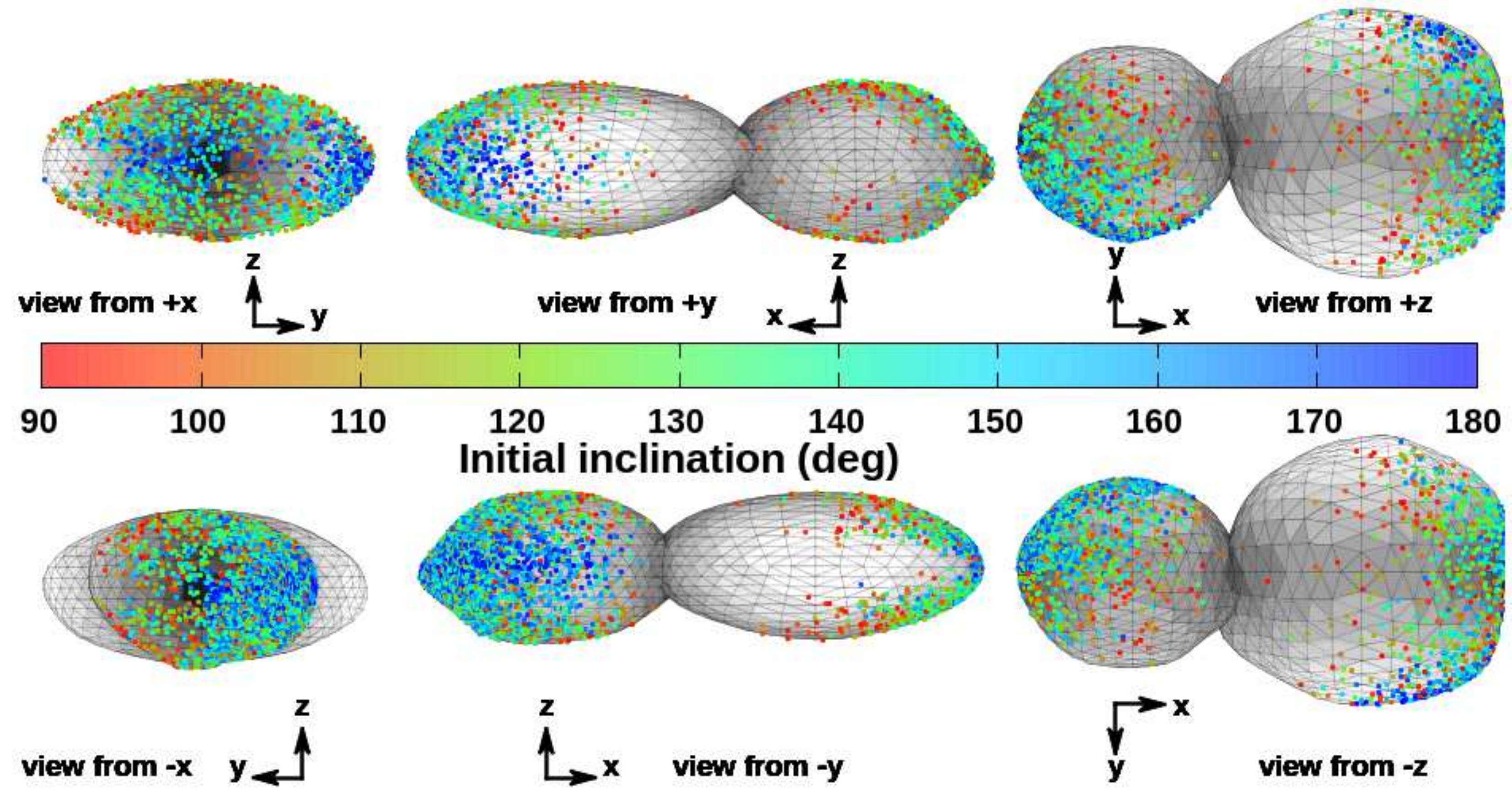}}
  \caption{Map of distribution of impacts over Arrokoth's surface. Retrograde impact sites are denoted by the dots, where their colours represent the initial inclination of the particle orbit in the inertial frame. The geometric height of the surface of Arrokoth is computed by the colour gray code. These are numerical results for the simulated particles with initial conditions of type (III).}
  \label{fig:fall_2}
\end{figure*}

The distribution of impacts on the surface of Arrokoth contact binary are shown in Figs. \ref{fig:fall_1} and \ref{fig:fall_2} for particles with prograde and retrograde orbits, respectively. These figures are made for six perspective views: $\pm x$, $\pm y$, and $\pm z$ axes.

At first, in general, we note that the most bombarded areas are located around the biggest and intermediate axes ends ($\pm x$ and $\pm y$, respectively), where the impact sites occur in the proximity of the region located at the equator of Arrokoth contact binary. There is also a noteworthy number of particles that impact the surface of the lobes close to their polar areas, where are situated the edges of the smallest axis ($\pm z$). 

We also note that from the total of prograde particles which collide with the large lobe, most of them ($\sim 75$\%) populate its eastern region ($+y$). For the small lobe, the situation is the opposite. From the total of prograde impacts across its surface, most of them ($\sim 64$\%) occurs preferentially in the western area ($-y$). This behavior also appears for the particles initially with retrograde orbits, except that the eastern region of the large lobe has only slightly more falls ($\sim 56$\%) than the western one ($\sim 44$\%), taking into account the total amount of collisions on its surface. Besides, the western region of the small lobe is predisposed to attach a higher number of particles ($\sim 85$\%) than the same region of the large lobe ($\sim 15$\%) (Tab. \ref{tab:fall_1}), considering the total number of retrograde falls on their surfaces. It is important to mention that the position of such impact areas for a rotating contact binary is quite different. It depends on the shape and the value of the rotation period of the minor body. This surface pattern of the impact points for prograde orbits is also expected for fast rotators, which have a dumb-bell shape of the same mass \citep{Vasilkova2003}. Here, we extend this result for prograde and retrograde trajectories around an irregular-shaped contact binary slow rotator, such as Arrokoth.

Table \ref{tab:fall_1} summarizes the percentages of total falls across the individual lobes for a spherical cloud initially orbiting Arrokoth. Note that there are significant differences between the distribution of impacts at the surface of each lobe for particles initially on the prograde and retrograde orbits. During the entire simulation time of $1.14$\,yr, almost half of the initial amount of prograde particles hits each lobe ($\sim 52$ for the large lobe and $\sim 48$ for the small lobe). Otherwise, particles in retrograde orbits impact the surface of the small lobe ($\sim 65$\%) more often than the large one ($\sim 35$\%).

Table \ref{tab:fall_2} gives the percentages of collisions, escapes, and remaining particles in the Arrokoth contact binary system as a function of radial distance from the barycentre of Arrokoth. We investigated ten different intervals of spherical cloud shells from initial conditions of type (III) integration, which were filled uniformly with 2,000 particles each one. We also explored particles that have initially prograde and retrograde orbits.
Table \ref{tab:fall_2} shows that for the spherical cloud shells very close to the surface of Arrokoth, the number of collisions with the surface of Arrokoth dominates the dynamics of the system with a maximum value between 17-20\,km, for prograde and retrograde trajectories. Far away from the surface of Arrokoth, the number of collisions reduces, and the number of escapes of the system rises until it reaches a maximum value between 32-35\,km for prograde orbits. The number of escapes for retrograde orbits reaches a maximum value earlier between 23-26\,km. There are no remaining prograde particles for initial radial distances $\leq 35$\,km. Meanwhile, retrograde particles start to survive along the integration for initial radial distances $\geq 20$\,km. Note that the equilibrium points around Arrokoth contact binary have radial distances from its barycentre in the approximate range between 20 and 23\,km (Tab. \ref{tab:math_3}). An initial radial distance $\geq 44$\,km indicates that all retrograde particles remain in the system for the entire simulation.
\begin{table}
 \centering
  \caption{Percentages of falls over the surface of the individual lobes of the Arrokoth contact binary for initial conditions of type (III) integration. Eastern and Western lines represent the percentages of falls, taking into account the total number of collisions on the surface of each lobe. The Total line indicates the falls over each lobe surface, considering the total amount of collisions across the entire surface of Arrokoth.}
 \label{tab:fall_1}
 \scalebox{1.1}
{
 \begin{tabular}{ccc}
  \toprule
    Region & Large & Small \\
  \hline
  \multicolumn{3}{c}{Prograde} \\
  Eastern ($+y$) & 75.2 & 35.7 \\
  Western ($-y$) & 24.8 & 64.3 \\
  Total & 52.4 & 47.6 \\
  \hline
  \multicolumn{3}{c}{Retrograde} \\
  Eastern ($+y$) & 55.5 & 15.3 \\
  Western ($-y$) & 44.5 & 84.7 \\
  Total & 35.4 & 64.6 \\
  \hline
 \end{tabular}}
 \end{table}
\begin{table}
 \centering
  \caption{Percentages of collisions, escapes, and remaining particles in the Arrokoth contact binary system as a function of radial distance from Arrokoth's barycentre for initial conditions of type (III) integration.}
 \label{tab:fall_2}
 \scalebox{1.0}
{
 \begin{tabular}{cccc}
  \toprule
    Distance (km) & Collisions & Escapes & Remaining \\
  \hline
  \multicolumn{4}{c}{Prograde} \\
  17-20 & 76.6 & 23.4 & 0 \\
  20-23 & 74.1 & 25.9 & 0 \\
  23-26 & 68.7 & 31.3 & 0 \\
  26-29 & 51.6 & 48.4 & 0 \\
  29-32 & 29.5 & 70.5 & 0 \\
  32-35 & 10.2 & 89.8 & 0 \\
  35-38 & 6.5 & 89.1 & 4.4 \\
  38-41 & 2.0 & 69.7 & 28.3 \\
  41-44 & 0.1 & 35.8 & 64.1 \\
  44-47 & 0 & 16.4 & 83.6 \\
  \hline
  \multicolumn{4}{c}{Retrograde} \\
  17-20 & 82.2 & 17.8 & 0 \\
  20-23 & 72.4 & 27.5 & 0.1 \\
  23-26 & 40.6 & 53.3 & 6.1 \\
  26-29 & 21 & 52.8 & 26.2 \\
  29-32 & 3.7 & 47.5 & 48.8 \\
  32-35 & 0.6 & 33 & 66.4 \\
  35-38 & 0.1 & 10.9 & 89 \\
  38-41 & 0 & 2.6 & 97.4 \\
  41-44 & 0 & 0.2 & 99.8 \\
  44-47 & 0 & 0 & 100 \\
  \hline
 \end{tabular}}
 \end{table}

In general, it is well known that retrograde orbits are more stable than prograde ones (see for instance \citet{Hamilton1991,Ernesto2006}). Thus, a particle in a retrograde orbit is less affected by perturbing effects from the irregular geopotential of the Arrokoth contact binary. This feature is explained by the particle time length in the proximity of prominence at the surface of each lobe \citep{Amarante2021}. Furthermore, $\sim 32$\% of the initial sample of particles in prograde orbits hit the surface of Arrokoth. Otherwise, $\sim 23$\% of retrograde particles initially distributed around Arrokoth collide with its surface.

\begin{figure}
  \centering
  \fbox{\includegraphics[width=8.44cm]{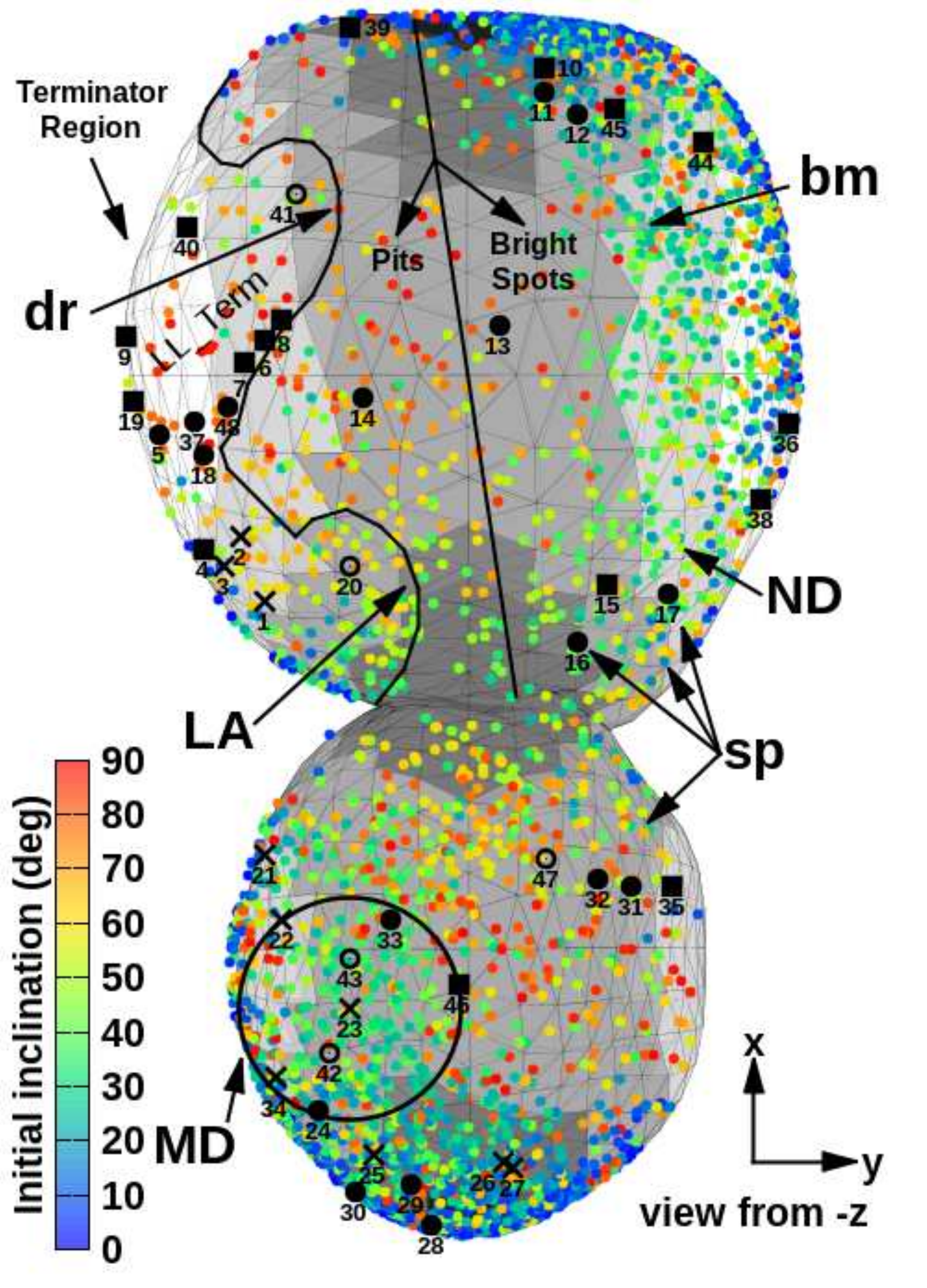}}
  \caption{Map of distribution of impacts across Arrokoth's surface from \textit{New Horizons}' spacecraft perspective of view. Prograde impact sites are represented by the dots, where their colours indicate the initial inclination of the particle orbit in the inertial frame. The geometric height is given by the colour gray code with the approximate locations of all known Arrokoth surface features defined in Fig. \ref{fig:math_2b}. These are numerical results for the simulated particles with initial conditions of type (III).}
  \label{fig:fall_3}
\end{figure}

Note from Figs. \ref{fig:fall_1} and \ref{fig:fall_2} that there are agglomerations of particles and empty areas along the entire equator of the surface of Arrokoth. Therefore, the distribution of impacts is not evenly around the equatorial region of Arrokoth. For example, the western and eastern areas of the large and small lobes, respectively, are those more depleted of falls for prograde orbits. The perspective of views $\pm y$ from Fig. \ref{fig:fall_1} indicate regions for prograde orbits that none of the falls occur. Most of these areas are in the eastern region of the small lobe and the western area of the large one. For retrograde trajectories, they also appear in the previous regions, like for the prograde ones. However, they cover a larger area (perspective of views $\pm y$ from Fig. \ref{fig:fall_2}). Furthermore, in the right-eastern region of the large lobe, there are sites with empty falls, as well as in its left-northern and southern areas (perspective of views $\pm z$ from Fig. \ref{fig:fall_2}). These locations indicate that the impacts accumulate preferentially near the low-mid-altitudes close to longitudes of MD impact crater and Bright spots area (Fig. \ref{fig:fall_3}). 

Figure \ref{fig:fall_3} shows the general picture for prograde falls on the surface of Arrokoth contact binary as seen by \textit{New Horizons} space probe's close-approach. We zoomed in the perspective view $-z$ of Fig. \ref{fig:fall_1} from initial conditions of type (III) integration. The surface features with their approximate locations were previously defined in Fig. \ref{fig:math_2b}. Note that the falls cover almost all the surface of Arrokoth. They spread across the high altitudes of the large lobe's surface and concentrate over the surface of the small lobe near the MD crater. Even a few of them reach the neck. However, there is a preference for accumulation of particles in the equatorial region close to the MD crater and Bright spots area. The final outcome of flux of particles are placed at the highest values of the geometric height on the eastern region of the large lobe, where are located the Bright spots area (right-hand side of Fig. \ref{fig:to_3}). In the regions near a valley, like the MD impact crater, the gravitational force from the irregular binary geopotential of Arrokoth makes a particle hit its surface early after the crater boundary. So, depending on the trajectory of a particle (prograde or retrograde), the fall will occur inside or outside the crater. Thus, this dynamic mechanism is responsible for retaining a considerable number of particles in the environment near the MD crater. This effect was seen before, for example, around the peaks across asteroid Bennu's surface \citep{Amarante2021}. The binary power-gravity feature of Arrokoth could be used to understand this effect, as discussed in the forthcoming paragraphs. Apart from that, surface features around the boundary of combined geologic units (LL\_Term) of the large lobe and in the eastern equatorial area of the small one are those more empty of falls.

Figures \ref{fig:fall_1}, \ref{fig:fall_2} and \ref{fig:fall_3} also show that the distribution of impacts over the surface of Arrokoth contact binary is quite dependent on the particle initial inclination. The particle trajectories initially with low inclinations are related to the accumulation of impacts along the equator of the Arrokoth contact binary. Nevertheless, particle orbits initially with high inclinations are more likely to impact the surface regions of the lobes near their poles.
Most particles in prograde orbits have impact sites between latitudes -20$^\circ$ and 20$^\circ$. They account for average inclinations of $35.5^\circ$ in the middle-latitude areas of the Arrokoth's surface. They cover $75$\% of the total number of impacts. Hight altitudes are those more empty of falls, with only $3$\% of the initial number of prograde falls. This same pattern is similar for retrograde orbits. Beyond that, most of the particles that impact the surface of Arrokoth in prograde trajectories fall in longitude intervals [0$^\circ$,60$^\circ$] and [180$^\circ$,240$^\circ$].

\begin{figure*}
  \centering
  \fbox{\includegraphics[width=15cm]{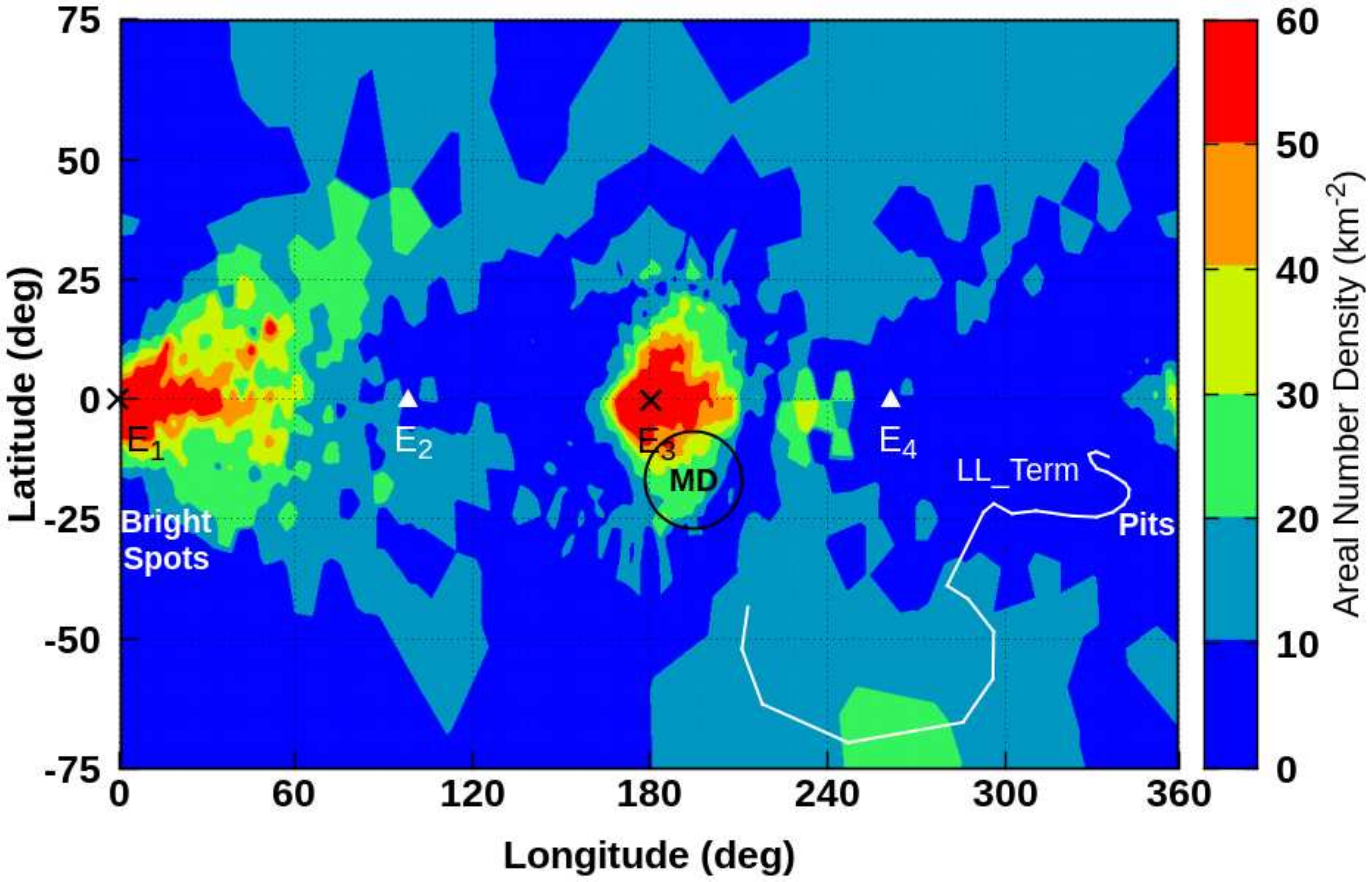}}
  \caption{Map of the areal number density across the surface of Arrokoth contact binary for particles that impact its surface in prograde trajectories. The areal number density of impacts (colour box) is computed by summing the number of impact sites over each triangular face and then normalized to the corresponding triangular area. The equilibrium point projections across Arrokoth's surface are represented by X-cross and triangle marks. Some topographic features of Arrokoth are also shown in this plot.}
  \label{fig:fall_5}
\end{figure*}

The areal number density map is made in Fig. \ref{fig:fall_5} for the particles in prograde orbits. The density of impacts over the surface of the Arrokoth contact binary is calculated by summing the number of impact sites into each triangular face of the polyhedral surface mesh. After that, we normalize the sum of impacts by the area of the corresponding triangular face. In this figure, we also show the location of the equilibrium point projections and their topological stabilities.
Note that the equilibrium points E$_1$ and E$_3$ are across regions close to Bright spots area and MD impact crater, respectively. Such areas show a high number of particles that impact the surface of Arrokoth (reddish areas). Thus, the dynamical environment of odd-numbered (HU) equilibrium points could be through as a mechanism to allow the particle flux to collide with the Arrokoth's surface. This result corroborates with analysis on the bottom-left side of Fig. \ref{fig:lod_1}. From this figure, the local disk around equilibrium point E$_3$ is faster to shrink and delivers most of its particles in the equatorial area close to the MD impact crater (see animated Movie 2 available online). On the other hand, particles spend more time surrounding equilibrium points E$_2$ and E$_4$ (CU) before hit the surface, which shows a low number of collisions near their projections on the areal number density map (bluish areas).
\begin{figure*}
  \centering
  \includegraphics[width=8.44cm]{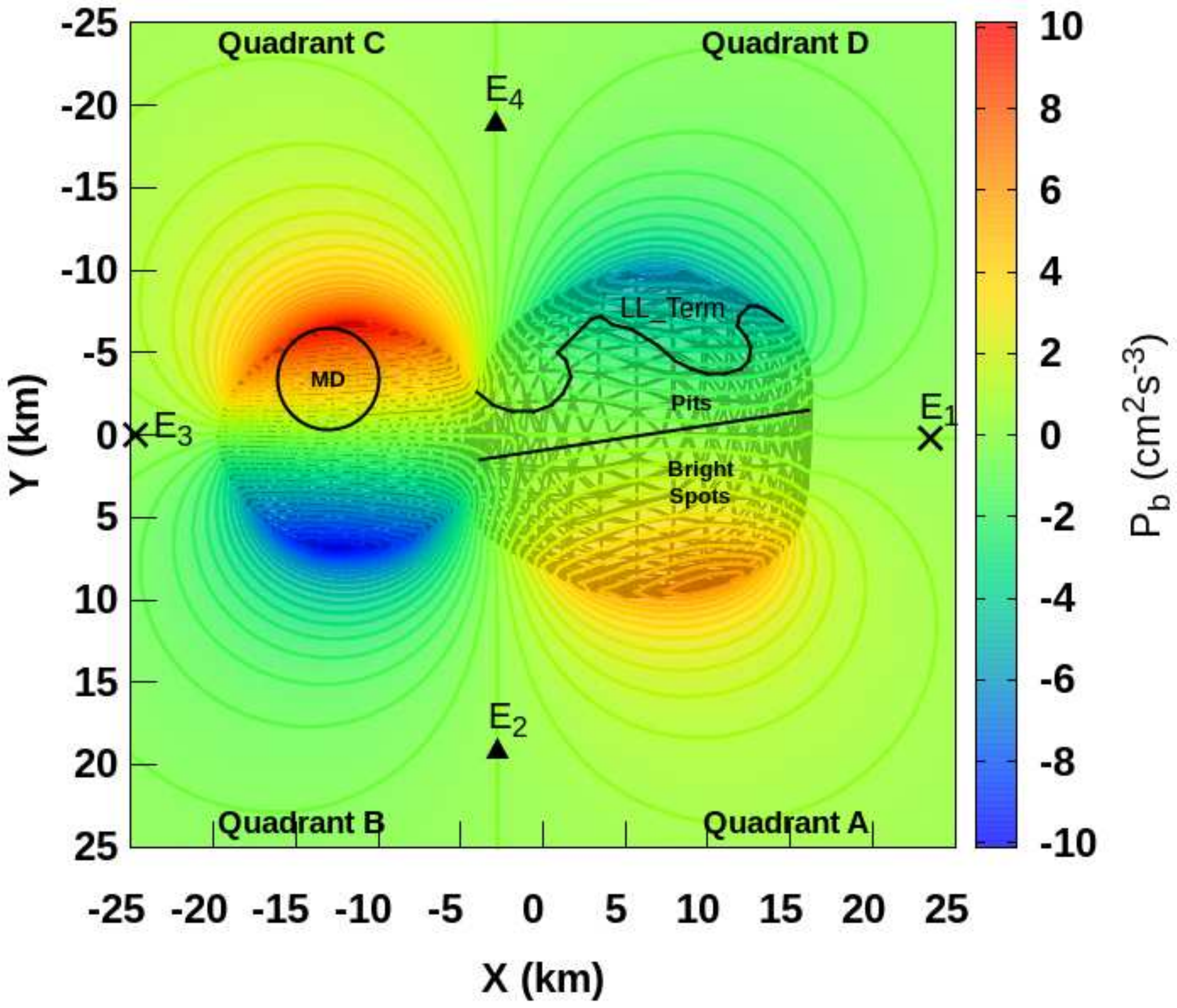}
  \includegraphics[width=8.44cm]{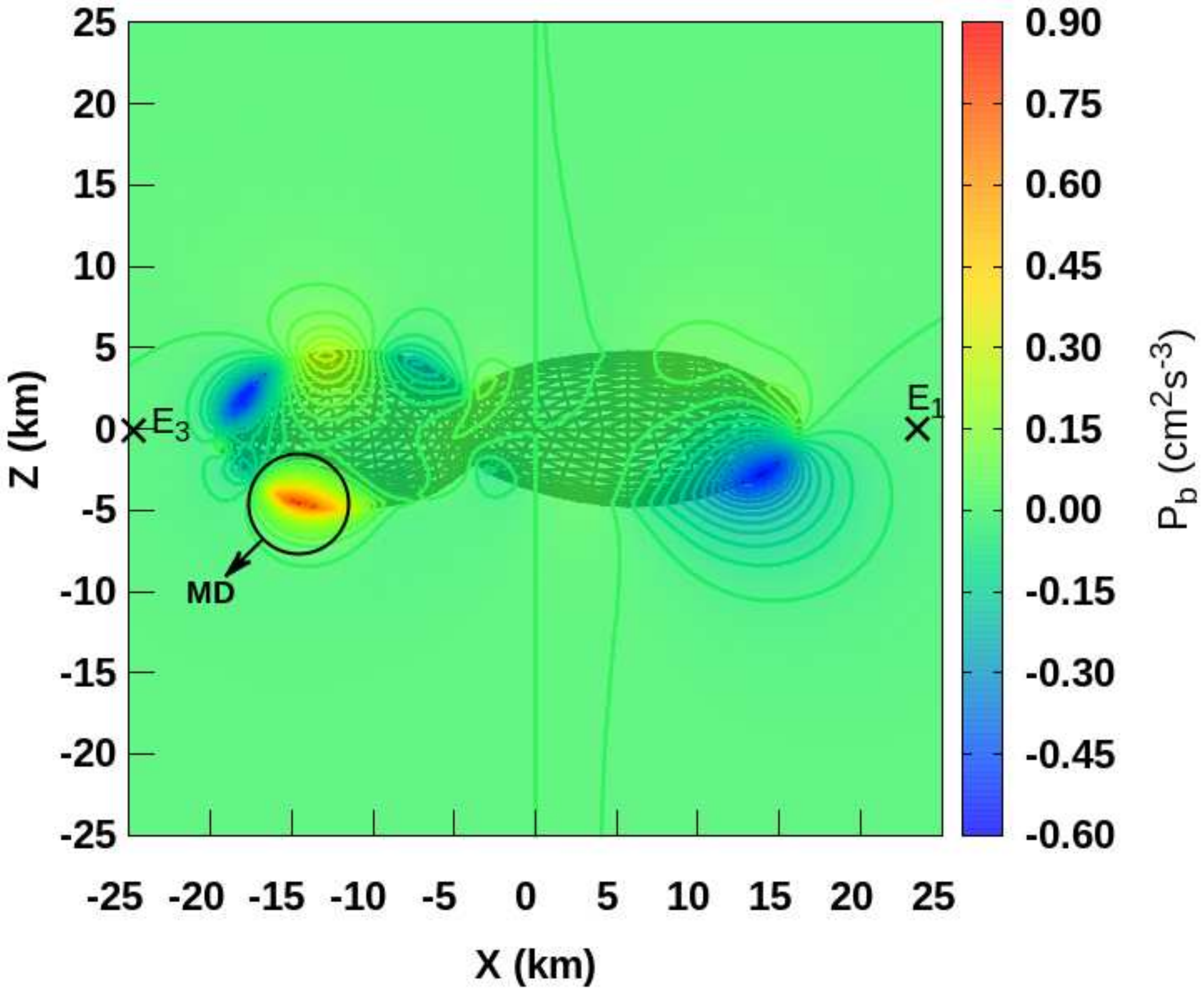}
  \caption{Binary Gravity-Power map for Arrokoth using its most recent 3-D polyhedral shape model \citep{Grundy2020,Spencer2020}. (left-hand side) The binary gravity-power map $P_b$ in the equatorial plane $xO-y$ as seen by \textit{New Horizons}' spacecraft. (right-hand side) The binary gravity-power contour levels in the projection plane $xOz$. The colour panels indicate the values of $P_b$, in cm$^2$\,s$^{-3}$. Shadowed areas sketch the shape of Arrkoth.}
  \label{fig:fall_6}
\end{figure*}

Using the spin rate vector $\pmb{\Omega}$ of Arrokoth and the position vector \textbf{r} of a particle, the binary gravity-power can be expressed as \citep{Yu2013,Amarante2020}:
\begin{align}
P_b & = -(\pmb{\Omega}\times\textbf{r})\cdot\nabla U_b.
\label{eq:fall_1}
\end{align}
Thus, Eq. \eqref{eq:fall_1} makes the binary gravitational force potential $U_b(\textbf{r})$ (Eq. \eqref{eq:math_1}) to be completly computed by the irregular geopotential of each lobe. A great advantage of Eq. \eqref{eq:fall_1} is to use its position-dependence to measure the enhancing and receding orbital energy of the particles in the flux around the lobes.

Figure \ref{fig:fall_6} express the behavior of gravity-power equation $P_b$ around Arrokoth contact binary for projections planes $xO-y$ (left-hand side) and $xOz$ (right-hand side). The left-hand side of Fig. \ref{fig:fall_6} shows the $P_b$ around the large and small lobes as seen by \textit{New Horizons} space probe's flyby. Note that there are four distinguishable regions where the particles experience-enhancing (Quadrants A and C) and receding (Quadrants B and D) in their orbital energy. On the eastern region of the surface of the large lobe (Quadrant A), $P_b$ is greater for the particle flux. In this location are placed the Bright spots area. Also, on the western area of the small lobe (Quadrant C), $P_b$ increases significantly close to the MD impact crater (red areas). In contrast, particles decrease their orbital energy close to the LL\_Term region of the large lobe (Quadrant D) and in the eastern region of the small one (blue areas). In addition, the right-hand side of Fig. \ref{fig:fall_6} suggests that the binary gravity-power field around the MD impact crater has the extreme values for projection plane $xOz$. In summary, the orbital energy of the particles in the flux around the lobes splits into complementary Quadrants, where $P_b$ is high and low. This effect is related to the outcome of particle flux (right-hand side of Fig. \ref{fig:to_3}), which also spreads in complementary regions over the surface of the lobes.

It is relevant to mention that the particle locations of impact areas across the surface of the Arrokoth contact binary are highly dependent on the adopted bulk density. The bulk density of the Arrokoth contact binary is largely unknown \citep{Spencer2020}. Depending on the considered bulk density for Arrokoth, the location and stability of the equilibrium points around it could change dramatically \citep{Amarante2020}. Since the phase space around Arrokoth is highly dependent on the equilibria, the assumed bulk density value could make the rotational Roche lobe region shrink or expand around it. In that way, particles will have different energy levels thresholds. Hence, they will follow the particle flux for a specific density value. Therefore, the dynamic behaviors of impacted particles over Arrokoth's surface could change because of the difference in the bulk density. This behavior is also related to surface stability. The slope could modify considerably over the surface of Arrokoth depending on the bulk density. The flow tendency of surface particles is to migrate towards the locations where the dynamic slopes have the lowest intensities, as in MD impact crater \citep{Amarante2020}. Then, depending on the adopted density value, the slopes will change over the surface characterizing different locations for the observed geological features of Arrokoth through the binary-gravity power thresholds.

\section{Final Remarks}
\label{sec:final}
This work provided outcomes about the fate of particles in the dynamical environment around the surface of \textit{New Horizons}' targeted Kuiper Belt object (486958) Arrokoth contact binary. For this purpose, we rebuilt the most recent 3-D polyhedral shape model of Arrokoth taking into account its high rotation period and approximate locations of all known topographic features in the same perspective seen by \textit{New Horizons}' spacecraft at close encounter time. The binary geopotential of Arrokoth was computed precisely for the dynamical model through the mascons technique.

We numerically explored the behaviors of the stability and evolution of simulated particles that impact the surfaces of the large and small lobes through the linear stability of the actual equilibrium points. For that, we recomputed the locations of the equilibria for the Arrokoth contact binary using its current shape model. Following, we made several samples of numerical experiments of simulated particles to investigate their stability and evolution in the environment around Arrokoth, considering the location and topological structure of the equilibrium points in the particles' fate. We have chosen three types of initial conditions for these purposes: (I) local disks, (II) ring of particles, and (III) spherical cloud.

For the initial conditions of type (I) integration, almost half of the initial particle amount from all local disks are removed in a few hours. The particles initially situated inside the ridge line around equilibrium points E$_1$--E$_3$ are the fastest to impact the surface of Arrokoth. In contrast, particles initially located inside the ridge line surrounding equilibrium points E$_2$--E$_4$ stay along with integration for intermediate and long times before they hit its surface or escape from the system. Apart from that, the local disks surrounding equilibrium points E$_3$ and E$_1$ deliver particles very quickly near the equatorial areas of the MD impact crater and Bright spots, respectively. Also, particles are placed at the location of the equilibrium points around Arrokoth in synchronous orbits. For this type of initial condition, the particle that departs from equilibrium point E$_3$ has the lowest lifetime before hitting Arrokoth's surface. The particle from equilibrium point E$_1$ is the next to leave its equilibrium point's neighborhood. The particles initially located at equilibrium points E$_4$ and E$_2$ have approximately the same lifetime, and they are those to remain at last in the numerical simulation.

The ring-type (II) was spread around Arrokoth's surface in a few hours. Most of the particles collided with the surface of the Arrokoth contact binary in $\leq 50$\,h. Half of them in the first $9$\,h. They account for $75$\% of the initial number of particles. Beyond that, escapes cover $25$\% of the pruned particles from the system. They start to escape only after $16$\,h. Despite, particles located near the longitudinal regions of equilibrium points E$_1$--E$_4$ and E$_3$--E$_2$ stay in the system for long times ($> 50$\,h).
Surface impact speeds show that particles will fall to the surface of Arrokoth with speeds $\leq 3$\,m/s. Most particles collided with Arrokoth's surface with speeds between 1.5-2.5\,m/s. This fact indicates that speeds $>3$\,m/s are more likely to tend particles escape from the Arrokoth's environment.
The map of the flux of particles around the surface of Arrokoth suggests that the preferred region for the particles remain along the integration is around equilibrium point E$_1$.
Considering the last 30\,min of the simulations, thus most of the particles inside the rotational Roche lobe will fall over Arrokoth's surface close to the longitudinal MD impact crater site of the small lobe or in the proximity to the region diametrically opposite to it, at the edges of Bright Spots area. On the other hand, a few of them will fall nearby the longitudinal LL\_Term area of the large lobe.
The SRP perturbation was also included in this type of numerical simulations for different sizes of particles.
The irregular binary gravity field of Arrokoth dominates particles larger than a couple of microns.

The last type of initial condition is the spherical cloud of particles (III). It was made to measure the accumulation of particles onto the surface of the lobes. The results associated with the spherical cloud suggest that most prograde particles, which collide with the surface of the large lobe, populate $75$\% of its eastern region. The small lobe has the opposite situation, with $64$\% of the particles falling over its western area from the total impact amount across its surface.
This behavior is quite different for the particles initially with retrograde trajectories. The large lobe attaches a few more than $50$\% of the total collisions on its eastern region. Meanwhile, a considerable number of $85$\% of the falls impact the western area of the small lobe.
$32$\% of the initial sample of particles in prograde orbits impacts Arrokoth's surface over the length of time of $1.14$\,yr. In contrast, particles in retrograde orbits are more likely to be stable. They account for $23$\% of the falls that impacted the surface of Arrokoth contact binary in the entire integration time.
The prograde particles start to remain in the system only for distances greater than 35\,km from the barycentre of Arrokoth. The distance for remaining particles in retrograde orbits is very close to the surface of Arrokoth ($\geq 20$\,km). Beyond that, more than $50\%$ of the total sample of prograde particles in a spherical cloud shell survive for the entire simulation for distances $\geq 41$\,km. All retrograde particles remain in the dynamical environment of Arrokoth for distances greater than 44\,km.

Finally, the behavior of the distribution of impacts across Arrokoth's surface is quite dependent on the particle initial orbital inclination and the adopted bulk density. The impact of particles in the equatorial area of the lobes is not uniformly distributed, with a preference for orbits of particles in the inertial frame that have low initial inclinations. In summary, the areal number density map shows that the distribution of impacts accumulates preferentially into low-mid-altitudes close to the longitudes of MD impact crater and Bright spots area. On the other hand, a few of particles will impact the surface in the vicinity of the region diametrically opposite to them, as in the LL\_Term boundary. These results show the preferred locations for the fate of the particles over the Arrokoth's surface features for the assumed bulk density. They also relate the dynamical connection with the environment around the surface of the Arrokoth contact binary.

 \acknowledgments
The authors thank the Improvement Coordination Higher Education Personnel - Brazil (CAPES) - Financing Code 001 and National Council for Scientific and Technological Development (CNPq, proc. 305210/2018-1). This research was financed in part by the thematic project FAPESP (proc. 2016/24561-0), and it also had computational resources provided by the Center for Mathematical Sciences Applied to Industry (CeMEAI), funded by FAPESP (grant 2013/07375-0). We are also grateful to the entire \textit{New Horizons} mission team for making the encounter with KBO Arrokoth possible.

 \begin{dataavailability}
 The data underlying this manuscript will be shared upon reasonable request to the corresponding author.
 \end{dataavailability}

\begin{codeavailability}
 Simulation codes used to generate the results are available online at \url{https://github.com/a-amarante}.
 \end{codeavailability}

 \begin{orcid}
A. Amarante \href{https://orcid.org/0000-0002-9448-141X}{\includegraphics[scale=0.5]{orcid_16x16.pdf}} \url{https://orcid.org/0000-0002-9448-141X}\\
O. C. Winter \href{https://orcid.org/0000-0002-4901-3289}{\includegraphics[scale=0.5]{orcid_16x16.pdf}} \url{https://orcid.org/0000-0002-4901-3289}
 \end{orcid}

 \begin{ethics}
 \begin{conflict}
The authors declare no conflict of interest.
 \end{conflict}
 \end{ethics}


%
 \bibliographystyle{Amarante_spr-mp-nameyear-cnd}  
 \bibliography{Amarante_biblio-u1}                

%

\appendix

\section{Collisional and Escape Criteria}
\label{sec:minor-mercury}

The inertia tensor of the Arrokoth contact binary can be used to measure the dimensions of an `equivalent' triaxial ellipsoid around Arrokoth's surface \citep{Dobrovolskis1996}. This ellipsoid has the following dimensions with principal semi-axes $a$, $b$, and $c$ computed by:
\begin{eqnarray}
\label{eq:app_1}
a=\sqrt{\frac{5(I_{yy}+I_{zz}-I_{xx})}{2M}}, \nonumber \\
b=\sqrt{\frac{5(I_{xx}+I_{zz}-I_{yy})}{2M}},           \\
c=\sqrt{\frac{5(I_{xx}+I_{yy}-I_{zz})}{2M}}. \nonumber
\end{eqnarray}
The most recent shape model of Arrokoth contact binary from \citet{Spencer2020} applied in Eqs. (\ref{eq:app_1}) leads us to consider this peculiar minor body with principal semi-axes of $\sim 21 \times 9 \times 5$\,km.
Our collisional code is implemented considering that the particle hits the surface of Arrokoth when it passes through its polyhedric triangular mesh. The algorithm starts to verify if the particle impacted Arrokoth's surface when its trajectory is inside the equivalent ellipsoid. This mechanism is performed to reduce the computational time effort during the integration. Besides that, the code uses the ray-casting method \citep{ROTH1982109} to find the approximate site across all triangular faces that have been impacted by the particle. The signals of the determinants from fives tetrahedrons are used for this purpose:

$\left| \begin{array}{cccc}
x_0 & y_0 & z_0 & 1 \\ 
x_1 & y_1 & z_1 & 1 \\ 
x_2 & y_2 & z_2 & 1 \\ 
x_3 & y_3 & z_3 & 1
\end{array} \right|$,
$\left| \begin{array}{cccc}
x & y & z & 1 \\ 
x_1 & y_1 & z_1 & 1 \\ 
x_2 & y_2 & z_2 & 1 \\ 
x_3 & y_3 & z_3 & 1
\end{array} \right|$,
$\left| \begin{array}{cccc}
x_0 & y_0 & z_0 & 1 \\ 
x & y & z & 1 \\ 
x_2 & y_2 & z_2 & 1 \\ 
x_3 & y_3 & z_3 & 1
\end{array} \right|$,
$\left| \begin{array}{cccc}
x_0 & y_0 & z_0 & 1 \\ 
x_1 & y_1 & z_1 & 1 \\ 
x & y & z & 1 \\ 
x_3 & y_3 & z_3 & 1
\end{array} \right|$,
$\left| \begin{array}{cccc}
x_0 & y_0 & z_0 & 1 \\ 
x_1 & y_1 & z_1 & 1 \\ 
x_2 & y_2 & z_2 & 1 \\ 
x & y & z & 1
\end{array} \right|$,

\noindent where $(x, y, z)$ is the particle coordinates. $(x_0, y_0, z_0)$ (polyhedron's barycentre), $(x_1, y_1, z_1)$, $(x_2, y_2, z_2)$, and $(x_3, y_3, z_3)$ are the tetrahedron vertices. Therefore, the particle impact the surface of Arrokoth contact binary if the previous five determinants have the same sign.

Additionally, our code also uses escape criteria as a periodic effect. The particle is removed from the numerical simulation if its orbital radius is outside a sphere, with a 100\,km radius, far away from the dynamical environment around Arrokoth. Both criteria are implemented into Minor-Mercury package\footnote{\url{https://github.com/a-amarante/minor-mercury}.} \citep{Amarante2021}, an N-body integrator from the original Mercury code \citep{Chambers1999} to handle with an irregular-shaped minor body.

\end{document}